\documentclass[tighten, twocolumn, trackchanges, 12pt]{aastex62}
\def\apjl{{ApJ}}                
\def\apjs{{ApJS}}               
\def\ao{{Appl.~Opt.}}           
\def\aap{{A\&A}}                

\usepackage{listings}
\usepackage{color}

\definecolor{dkgreen}{rgb}{0,0.6,0}
\definecolor{gray}{rgb}{0.5,0.5,0.5}
\definecolor{mauve}{rgb}{0.58,0,0.82}
\definecolor{golden}{rgb}{0.86,0.65,0.01}

\usepackage{graphicx}
\usepackage{hyperref}  
\begin{document}


\title[GaiaEDR3mock]{A Gaia early DR3 mock stellar catalog: Galactic prior and selection function}

\correspondingauthor{Jan Rybizki - rybizki@mpia.de}

\author[0000-0002-0993-6089]{Jan Rybizki}
\affil{Max Planck Institute for Astronomy,
	K\"onigstuhl 17, D-69117 Heidelberg, Germany}

\author{Markus Demleitner}
\affil{Astronomisches Rechen-Institut, Zentrum f{\"u}r Astronomie der Universit{\"a}t Heidelberg, M{\"o}nchhofstrasse 12-14, D-69120 Heidelberg, Germany}
\author{Coryn Bailer-Jones}
\affil{Max Planck Institute for Astronomy,
	K\"onigstuhl 17, D-69117 Heidelberg, Germany}
\author{Piero Dal Tio}
\affil{Osservatorio Astronomico di Padova -- INAF, Vicolo dell’Osservatorio 5, I-35122 Padova, Italy}
\affil{Dipartimento di Fisica e Astronomia Galileo Galilei, Universit\`a di Padova, Vicolo dell'Osservatorio 3, I-35122 Padova, Italy}
\author[0000-0001-8726-2588]{Tristan Cantat-Gaudin}
\affil{Institut de Ci\`encies del Cosmos, Universitat de Barcelona (IEEC-UB), Mart\'i i Franqu\`es 1, E-08028 Barcelona, Spain}
\author[0000-0001-9256-5516]{Morgan Fouesneau}
\affil{Max Planck Institute for Astronomy,
	K\"onigstuhl 17, D-69117 Heidelberg, Germany}
\author{Yang Chen}
\affil{Dipartimento di Fisica e Astronomia Galileo Galilei, Universit\`a di Padova, Vicolo dell'Osservatorio 3, I-35122 Padova, Italy}
\author[0000-0002-6274-6612]{Ren\'{e} Andrae}
\affil{Max Planck Institute for Astronomy,
	K\"onigstuhl 17, D-69117 Heidelberg, Germany}
\author{L\'eo Girardi}
\affil{Osservatorio Astronomico di Padova -- INAF, Vicolo dell’Osservatorio 5, I-35122 Padova, Italy}

\author{Sanjib Sharma}
\affil{Sydney Institute for Astronomy, School of Physics, The University of Sydney, NSW 2006, Australia}
\affiliation{ARC Centre of Excellence for All Sky Astrophysics in Three Dimensions (ASTRO-3D)}


\vspace{10pt}

\begin{abstract}

We present a mock stellar catalog, matching in volume, depth and data model the content of the planned Gaia early data release 3 (Gaia EDR3). We have generated our catalog (GeDR3mock) using \texttt{galaxia}, a tool to sample stars from an underlying Milky Way (MW) model or from N-body data. We used an updated Besan\c{c}on Galactic model together with the latest PARSEC stellar evolutionary tracks, now also including white dwarfs. We added the Magellanic clouds and realistic open clusters with internal rotation. We empirically modelled uncertainties based on Gaia DR2 (GDR2) and scaled them according to the longer baseline in Gaia EDR3. The apparent magnitudes were reddened according to a new selection of 3D extinction maps. 

To help with the Gaia selection function we provide all-sky magnitude limit maps in G and BP for a few relevant GDR2 subsets together with the routines to produce these maps for user-defined subsets. We supplement the catalog with photometry and extinctions in non-Gaia bands. The catalog is available in the Virtual Observatory\footnote{\url{http://dc.g-vo.org/tableinfo/gedr3mock.main}} and can be queried just like the actual Gaia EDR3 will be. We highlight a few capabilities of the Astronomy Data Query Language (ADQL) with educative catalog queries. We use the data extracted from those queries to compare GeDR3mock to GDR2, which emphasises the importance of adding observational noise to the mock data. Since the underlying truth, e.g. stellar parameters, is know in GeDR3mock, it can be used to construct priors as well as mock data tests for parameter estimation. 

All code, models and data used to produce GeDR3mock are linked and contained in \texttt{galaxia\_wrap}\footnote{\url{https://github.com/jan-rybizki/Galaxia_wrap}}, a python package, representing a fast galactic forward model, able to project MW models and N-body data into realistic Gaia observables.

\end{abstract}
%
%
%
%
%

\section{Introduction}
The Gaia mission \citep{2016A&A...595A...1G} and second Data Release 2 (GDR2)
\citep{2018A&A...616A...1G}
have provided positions, parallaxes, proper motions, and three photometric bands for 1.3\,billion sources across the sky. It also provided effective temperatures, luminosities, extinctions,  and radial velocities for various subsets of these sources. While this has led to an unprecedented rich view of our Milky Way system \citep[i.a.][]{2018Natur.563...85H,2018A&A...618A..93C,2018MNRAS.478..611B}, it is at the same time hard to understand the limits of this data set. To help with this, the community has produced mock stellar catalogs that have similar selections to, and provide the same observables, as Gaia, in which the underlying truth is known. \citet{2018MNRAS.481.1726G,2020ApJS..246....6S}  used  N-body cosmological simulations of Milky Way-like galaxies. These have been used to interpret patterns in the stellar phase-space structure seen in GDR2 in terms of our Galaxy's merger history \citep{2019arXiv190904679B, 2020arXiv200106009G}, and they have been used to estimate the mass of our Galaxy \citep{2019MNRAS.487L..72G}. A slightly different approach was taken by some of the present authors in \citet{2018PASP..130g4101R}, were we used an underlying Milky Way model \citep{2003A&A...409..523R} to produce a mock stellar catalog with \texttt{galaxia} \citep{2011ApJ...730....3S}, a tool to sample stars from density distributions or N-body data. We published this in the same way as GDR2, namely via \texttt{ADQL} and mimicking the GDR2 data model. This proved useful for testing the Gaia selection function \citep{2018A&A...616A..37B,2019ApJ...887..237C} and also to estimate false positive rates in common proper motion pairs \citep{2018MNRAS.480.4884E,2020ApJS..246....4T}. It served also as a Galaxy prior \citep{2018AJ....156...58B} and provided an easy way to query a Milky Way model \citep{2019ApJ...881..164Y,2020MNRAS.tmp...78A} or estimate starcounts for future surveys \citep{2019ApJ...883..107C}.

In this paper we present Gaia early DR3 mock (GeDR3mock),
a simulated Gaia catalog with entries for 1,573,457,319 individuals stars brighter than G\,=\,20.7\,mag. It is intended as a community service for the preparation of the upcoming Gaia Early Data Release 3 (Gaia EDR3).
 Compared to our GDR2mock catalog
\citet{2018PASP..130g4101R}, 
for GeDR3mock
we have updated the Milky Way model \citep{2014A&A...564A.102C} and have added the Magellanic Clouds and over 1,000 open clusters \citep{2018A&A...618A..93C,2013A&A...558A..53K}.
We simulate observational uncertainties empirically using GDR2 uncertainties scaled to the longer baseline of 34\,months for Gaia EDR3 (compared to 22\,months in GDR2). 
We again mimic the GDR2 data model, and additionally provide all underlying stellar parameters, e.g. teff, logg, feh, age, extinctions in all bands, initial and current mass, and which galactic component the star belongs to. All values provided in the catalog are noise free and we provide all values for all stars in the catalog, i.e.\ including those that will be absent from GeDR3 or even Gaia DR3. For example we provide radial velocities for all stars, which means that the user has to apply an appropriate selection. We assist with the selection function by providing both, maps of limiting magnitudes for four different GDR2 selections\footnote{The RVS sample is missing and will be addressed in a forthcoming publication.} (all sources, all with parallax, all with BP-RP, all with parallax and BP-RP),
and the tools we used to create the maps  \citep{2019ascl.soft01005R}. We provide example \texttt{ADQL} queries to illustrate how to access the data.

The paper is structured as follows: In Section\,\ref{sec:generation} we sketch the generation of GeDR3mock. Section\,\ref{sec:selection} discusses selection effects of the Gaia instrument, following by a comparison to GDR2 in Section\,\ref{sec:comparison}. In Section\,\ref{sec:catalog} we discuss the catalog content and limitations. We provide example queries in Section\,\ref{sec:example_queries}.

\section{Catalog generation}
\label{sec:generation}
Our catalog has been generated using \texttt{galaxia} \citep{2011ApJ...730....3S} a tool to turn an underlying chemo-dynamical Milky Way model via stellar isochrones into a synthetic or `mock' stellar catalog. It also has the functionality to turn N-body data into mock stellar particles, which we use to generate the Magellanic Clouds and open clusters. 
We use version 0.8.1 of 
 \texttt{galaxia} \citep{2019MNRAS.tmp.2471S} and made some modifications to the code that we explain below. We have linked the final version of our \texttt{galaxia} code in the \texttt{galaxia\_wrap}\footnote{\url{https://github.com/jan-rybizki/Galaxia_wrap}} python package. Both can be used in interplay to redo or customise our catalog.

\subsection{The Milky Way model}
The underlying Galaxy model of \texttt{galaxia} is based on the Besan\c con \citep{2003A&A...409..523R} model. Since 2003 the Besan\c con model has seen many updates for various Galactic components. We have implemented a selection of these changes and list them in the following subsections. For each Galactic component we indicate the population ID (\texttt{popid}) which can be used to only select stars of a specific component. Basic information on age and local mass normalisation of the thin- and thick-disk components can be inspected in Table\,\ref{tab:local_mass}.
\label{sec:galaxia}

\subsubsection{Thin disk - popid = 0-6}
We use a thin disk scale length of 2.2\,kpc
for popid 1 to 6
\citep{2009A&A...495..819R}, but use the fiducial 5\,kpc 
for the youngest disk population
(popid = 0) as in \citet{2014A&A...564A.102C}. The star formation rate (SFR) 
is modelled as exp(-0.12$\tau$), where $\tau$ is the time from 10\,Gyr ago, in accordance with \citet{2014A&A...564A.102C}\footnote{Though the SFR is still piecewise flat for each thin disk population, cf. \citet[][tab. 2]{2011ApJ...730....3S}}. We use the KH-v6 initial mass function (IMF) from \citet[tab.\,1]{2014A&A...564A.102C} for the thin disk. For the metallicity we implemented the values from \citet[tab.\,5]{2012A&A...543A.100R}.

\subsubsection{Thick disk - popid = 7}
For the thick disk, we implemented an age spread of 1\,Gyr \citep{2019MNRAS.tmp.2471S} and left the mean at 11\,Gyr. The thick disk metallicity is set to $-0.48\pm 0.3$ dex \citep{2014A&A...564A.102C}.

\subsubsection{halo - popid = 8}
We set the age of the halo to 13\,Gyr instead of the default 14\,Gyr because of isochrone limitations. The metallicity is $-1.5\pm0.5$ dex \citep[tab.\,5]{2012A&A...543A.100R}. The velocity dispersion is taken from \cite[tab.\,7]{2012A&A...543A.100R}.

\subsubsection{Bulge - popid = 9}
Metallicity is $0.0\pm 0.2$ \citep[tab.\,5]{2012A&A...543A.100R}.
Velocity dispersion is also taken from \citep[tab.\,7]{2012A&A...543A.100R}

\subsubsection{Magellanic clouds - popid = 10}
The most prominent extragalactic features in the sky density maps of GDR2 data are the Magellanic clouds (MCs) (see the middle panel of Figure\,\ref{fig:catalog_comparison}). We include a simple model of the MCs in order to study first order selection effects that occur in such dense regions at large distances. In case the user does not want to include the MCs when querying GeDR3mock, this can be done adding the following to a query:
\begin{lstlisting}
WHERE popid != 10 -- this is part of an ADQL query 
\end{lstlisting}
To generate N-body particles (from which \texttt{galaxia} then produces mock stellar particles) that represent the MCs, we use the parameters of the MC model tabulated in \citet[tab.\,10]{2012A&A...543A.100R} and assumed a constant star formation rate for both MCs. The sky position is taken from \citet{2003A&A...412...45P}. We arbitrarily set the velocity dispersions to 20 and 10 km/s for the LMC and SMC respectively. The stellar masses are set to $3.8\times10^9$\,M$_\odot$ for the LMC and to $6\times10^8$\,M$_\odot$ for the SMC, values that  reproduce approximately the starcounts in those regions of the sky. 
To build the LMC we drew
$10^5$ particles from a Gaussian distribution with a standard deviation of 1.075\,kpc.
We assume spherical symmetry for the SMC as well for which we drew $10^4$
particles with a standard deviation of 0.525\,kpc. The velocity distribution is randomly applied to each particle (relative to the 3D velocity of the centre-of-mass of the LMC given in \citet[tab.\,10]{2012A&A...543A.100R}) using a normal distribution and neglecting the position of the particle within the MCs (which we accept is dynamically inconsistent). This means that GeDR3mock MCs kinematics should not be used to compare to the detailed Gaia observations, which a.o. allows for rotation field inference in these galaxies. For \texttt{galaxia} to generate mock stellar particles from these $10^5$ and $10^4$ N-body particles we calculate a 6D smoothing length using EnBiD\footnote{\url{https://sourceforge.net/projects/enbid/}} \citep{2006MNRAS.373.1293S}.

\subsubsection{Open cluster - popid = 11}
Our underlying Milky Way model is smooth, but we know that the real Galaxy has many localised overdensities in phase-space, like moving groups, and open and globular cluster. Unraveling and cataloging such structures with the help of \textit{Gaia} data is an active topic of research \citep[e.g.][]{2019ApJS..245...32L,2020arXiv200107122C}. We add mock open star clusters to our catalog, so that the astronomical community can train their algorithms to detect them and to extract their underlying astrophysical parameters.


As an input catalog we use 1118 real clusters from \citet{2018A&A...618A..93C} and \citet{2013A&A...558A..53K}. We mock up the unknown astrophysical parameters of these in order to create an underlying truth, from which we can sample stars. The exact procedure can be inspected in notebook 7a of \texttt{galaxia\_wrap} (where also a fits file with the exact values can be found), but in brief, the procedure for assigning parameters to individual objects is as follows.
\begin{itemize}
    \item  The metallicity, [Fe/H], is forced to be 0.1 dex in the inner disk, -0.35 dex in the outer disk with a linear transition in between 8\,kpc and 12\,kpc galactocentric radius. We add Gaussian noise of $\pm$ 0.1 dex to these [Fe/H] values.
    \item Mass values of the clusters are picked from a truncated normal distribution (300 $<$ M$_\odot$ $<$ 2050, most clusters have low masses) and were sorted and assigned according to the number of member stars in GDR2. We chose the mass distribution in order to roughly reproduce the overall number of all cluster members, which is of order of 400K stars.
    \item We assume a solid body rotation for the stellar clusters with random spin axis. Rotational velocity is also correlated with the number of member stars (more stars mean higher rotational velocity). Velocities range from 0.1 to 0.7\,km/s and are given at the cluster radius.
    \item cluster center-of-mass positions and velocities were directly taken from the input catalog of 1118 clusters.
\end{itemize}
The mock data was generated using notebook 7. The particles contained in each cluster were distributed in a Plummer sphere using \texttt{amuse} \citep{2009NewA...14..369P}. To the resulting self-consistent velocity distribution we added the internal rotation depending on the position of each stellar particle with respect to the spin axis. The open cluster population can be easily queried\footnote{We report ADQL queries for didactic purposes and to help with reproducibility. They can be run on TAP services such as \texttt{topcat} \citep{2005ASPC..347...29T}, which also visualises the results. Alternatives include \texttt{PyVO} \citep{2014ascl.soft02004G,2019ASPC..521..483B} and web interfaces (e.g. \url{http://dc.zah.uni-heidelberg.de/__system__/adql/query/form}).} via:
\begin{lstlisting}
SELECT GAVO_NORMAL_RANDOM(pmra,pmra_error) AS pmra_obs, -- noise added values
GAVO_NORMAL_RANDOM(pmdec,pmdec_error) AS pmdec_obs 
FROM gedr3mock.main 
WHERE popid = 11 -- selects only open cluster stars
-- takes about 10 minutes
\end{lstlisting}
This query was used to generate the data for Fig.~\ref{fig:oc}, where observational noise has already been added via the \texttt{GAVO\_NORMAL\_RANDOM} function\footnote{It is not possible to reproduce the results of this function via e.g. a seed, due to the unspecified sequence in which the query results are returned.}. We show the proper motions for mock and GDR2 cluster members \citep{2020A&A...633A..99C} in orange and blue, respectively. Although the real clusters and their mock counterparts differ on a star-by-star basis, their statistical properties are (by design) in overall agreement.

Finding and characterising the mock clusters might be a good exercise to test the capabilities of detection methods to be used on the \textit{Gaia}~EDR3 data. If the user is not interested in those mock clusters, they can be excluded from a query via the statement:
\begin{lstlisting}
WHERE popid! = 11 -- de-selects the open clusters
\end{lstlisting}
\begin{figure}
	\includegraphics[width=\linewidth]{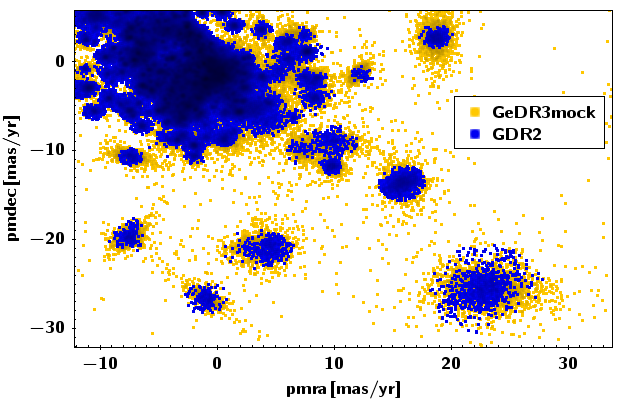}
	\caption{PMRA and PMDEC for open cluster member stars in GeDR3mock in orange and GDR2 in blue. Observational error is added to GeDR3mock from within ADQL.}
	\label{fig:oc}
\end{figure}

If users would like to mock up their own N-body data, e.g.\ clusters including tidal tails, streams or whole galaxies, they can adjust the procedure used in \texttt{galaxia\_wrap} notebook\,6\,\&\,7.

\subsubsection{Galactic warp and flare}
We update the parametrisation of the warp, based on \citet{1999ApJ...521..190G}, following \citet{2009A&A...495..819R}. Their comparison to 2MASS starcounts reveals that the displacement of the mid plane, characterised by the term $\gamma_\mathrm{warp}$ in the expression
\begin{equation}
    z_\mathrm{warp}(R) = \gamma_\mathrm{warp}\times (R-R_\mathrm{warp})\times \sin(\phi - \phi_\mathrm{warp})
\end{equation}

needs to be lowered from 0.18 to 0.09. $z_\mathrm{warp}(R)$ denotes the height of the warp above the plane. The starting galactocentric radius of the warp, $R_\mathrm{warp}$ is left at 8.4\,kpc. For the warp angle, $\phi_\mathrm{warp}$, we change the value from $0^\circ$ to $15^\circ$ in line with \citet{2004mim..proc..165Y}, which had no major effect on the fitting in \citet{2009A&A...495..819R}. 
\subsubsection{Thin- and thick-disk normalisation}
The various changes to the default \texttt{galaxia} MW model outlined above, especially to the SFR and IMF, result in a substantial change in the starcount distribution over all sky in our updated model. To gauge the new model to GDR2 data we produced models with different thin- and thick-disk normalisations, i.e. we rescaled the density distribution of the underlying model by a linear factor for thin- and thick-disk separately. We compared to local densities which are based upon \citet{1997ESASP.402..675J} data (c.f. Table\,\ref{tab:local_mass}) and global starcounts (c.f. Figure\,\ref{fig:catalog_comparison}). For the latter we applied HEALpix\footnote{\url{http://healpix.sf.net}} dependent G magnitude limits \citep{2018ascl.soft11018R} (as explained in Section\,\ref{sec:maglim}) to the mock and the real data and cut out the MCs. We also inspected how well the mock data would fit the real data, using a Poisson likelihood based on binned CMDs per HEALpix, where the HEALpix level is variable in order to have a similar amount of stars in each HEALpix (and therefore CMD). The exact procedure and algorithms can be looked up in \texttt{galaxia\_wrap} notebook 5\footnote{The computational time of a single all-sky evaluation is about 10\,minutes on a modern CPU, which makes it feasible to run inferences on multiple galactic parameters similar to \citet{2019MNRAS.tmp.2471S} or \citet{2019arXiv190404350P}. A somewhat superior approach in terms of computational cost can be found in \citet{2018A&A...620A..79M}.}.
A compromise between the overall starcounts, the local mass normalisation, and the CMD likelihood was then chosen by eye\footnote{Figures of the different metrics can be inspected in notebook 5a.} resulting in a thin disk normalisation of 0.9 and a thick disk normalisation of 0.8. The new thin disk normalisation of 0.9 applies to all thin disk populations\footnote{In \texttt{galaxia} the thin disk population 6 has been lowered by 20\,\%. We no longer apply this reduction.}, i.e. \texttt{popid}\,$\in[0,1,2,3,4,5,6]$.

\subsection{PARSEC-COLIBRI isochrones}\label{sec:isochrones}
The set of isochrones is the main astrophysical input that turns the underlying density distribution into mock stellar observations. Therefore we included the latest updates on these, as well as white dwarf tracks\footnote{White dwarfs were not included in GDR2mock}.
The basic isochrones come from \citet{2017ApJ...835...77M}, and are built joining the PARSEC evolutionary tracks from \citet{2012MNRAS.427..127B} with the thermally-pulsing asymptotic giant branch (AGB) from \citet{Pastorelli_2019}\footnote{\url{htts://stev.oapd.inaf.it/cmd}}. To these tracks, we add WD tracks from \citet{Bertolami_2016}, using cooling sequences of initial metallicity $Z=0.01$ from \citet{Renedo_2010}. These grids of white dwarfs have been extrapolated up to a final WD mass of $1.1\,M_\odot$ by using fitting relations. The derived isochrones are converted into the Gaia DR2 magnitudes by means of synthetic photometry performed with the YBC software \citep{2019A&A...632A.105C}\footnote{\url{htts://stev.oapd.inaf.it/YBC}}, in this case using the \citet{2018A&A...617A.138W} filter transmission curves for Gaia, which provides two BP bands, i.e. one for bright and one for faint magnitudes with a limit of $\mathrm{G}=10.87$\,mag.

For the generation of Gaia photometry \texttt{galaxia} uses the complete isochrone set. In order to calculate the extinction in all bands (Gaia, SDSS, 2MASS, UBV) and the photometry in other systems (SDSS, 2MASS, UBV) we use a gridded version of the isochrones, reducing the total number of model stars from 8,102,858 to 243,238. We create a grid with the following number of bins (step-size in parentheses) [boundaries in brackets]: [Fe/H] 36 (0.05\,dex)[-1.5,0.34]; $\log_{10}(T_{\mathrm{eff}})$ 162 (0.02) [2.45,5.68]; $\log_{10}(\mathrm{L_\odot})$ 217 (0.05) [-4.60, 6.24]. A combination of those three dimensions determines the \texttt{index\_parsec} (LLLTTTFFF, L = lum, T = teff, F = feh). The median of all stars that fall into a specific gridpoint is taken and also the standard deviation inspected\footnote{We exclude some TP-AGB stars which had extremely high extinction values resulting in strong outliers within our isochrone grid.}. We report here the 50 and 99 percentiles of the standard deviation for all bins of this grid for the other stellar parameters: log(age) [0.03,1.22]; initial mass [0.10,2.96]; current mass [0.03,4.14]; log(g) [0.04,0.51]; G [0.04,0.78]; G$_\mathrm{BP}$ [0.04,1.12]; G$_\mathrm{RP}$ [0.04,0.68]; G$_\mathrm{RVS}$ [0.04,0.63]. These photometric bands and extinctions can be queried via a separate table: \texttt{gedr3mock.parsec\_props}. An example is given in Section\,\ref{sec:parsec_props}. Due to the non-linear scaling of extinction with dust column density (reddening of an already reddened spectrum is weaker), extinction values are given for 6 different A0 values: 1,2,3,5,10,20\,mag. For an extinction law we used \citet{1989ApJ...345..245C} plus \citet{1994ApJ...422..158O}, with $R_V$=3.1, and higher order bolometric corrections have been taken into account. Links to the raw and reduced isochrone data are given in \texttt{galaxia\_wrap}. Notebooks on the generation of the grid can be found here\footnote{\url{https://github.com/jan-rybizki/Galaxia_wrap/tree/master/notebook/isochrone_generation}}.

\subsection{New 3D extinction map}
An integral part of a mock stellar catalog generation is the application of interstellar reddening due to dust, because most stars which would have a G magnitude brighter than 20.7 in the absence of dust have a fainter G magnitude after extinction has been added.
We build upon our experience with the \citet{2016ApJ...818..130B} combined dust map, which was put together using different 3D extinction maps (in order to get full-sky coverage). We replace the Bayestar 2015 map \citep{2015ApJ...810...25G} by Bayestar 2019 \citep{2019arXiv190502734G} up to $|b|<20^\circ$ and Bayestar 2017 \citet{2018MNRAS.478..651G} above (Bayestar 2017 has less clustering in low dust regions). Towards the Galactic center the combined map uses \citet{2006A&A...453..635M} which goes deeper since it is based on infrared data, whereas Bayestar requires photometric measurements in the optical as well. Parts that are not filled due to the Pan-STARRS \citep{2016arXiv161205560C} footprint are filled with \citet{2003A&A...409..205D}. See Figure 1 of \citet{2016ApJ...818..130B} for the footprint of each map. Resolution was increased to HEALpix level 9 (nside = 512, area of 47\,arcmin$^2$) from Healpix level 7 (nside = 128, 755\,arcmin$^2$) in GDR2mock and distance sampling is refined to 120 bins logarithmically sampled from 60\,pc to 60\,kpc\footnote{These values refer to Bayestar 2019. The other maps have different grids which are interpolated to that grid.}. The data cube is linked in \texttt{galaxia\_wrap} and methods for application are present in the \texttt{library/add\_extinction.py} file of \texttt{galaxia\_wrap}. $A_0$ (monochromatic extinction at lambda = 547.7nm in mag) values are interpolated linearly in distance while adjacent HEALpix values are not interpolated (HEALpix footprint is visible, as are the borders between the different extinction maps).

As reported before, the extinction in specific bands (G, BP\_bright, BP\_faint, RP and RVS; these two BP bands will be merged in a later step, described in Section\,\ref{sec:mag_cut}) has been precalculated for 6 different values of A$_0$: 1,2,3,5,10,20\,mag. In order to calculate the extinction in a specific band for a specific value of A$_0$ (that comes from the 3D extinction map), we make a cubic fit to those 6 values onto a finer grid between 0 and 20\,mag and then interpolate linearly to the exact A$_0$ (this two step interpolation is a compromise between accuracy and speed, when operating with large arrays of extinction).\footnote{To query A$_0$ values from the 3D positions of 10\,M sources and apply the band corrected extinction to each of the 5 bands takes about 1\,minute on a modern laptop.} For values of A$_0$ that are larger than 20\,mag, we linearly scale the value for A$_0=20$\,mag. The procedure outlined above illustrates the steps taken in \texttt{library/util.py:apply\_extinction\_curves()} and gives the extinction in the respective photometric band which we add to the apparent magnitude of the unreddened stars as generated by \texttt{galaxia}.

\subsection{Apparent magnitude cut}
\label{sec:mag_cut}
For GeDR3mock we compute all stars with G brighter than 20.7\,mag using \texttt{galaxia}.
Afterwards we apply our 3D extinction map and add absorption to each band. Thereafter we remove stars with G\,$>20.7$\,mag, which diminishes stars by a factor of 4. Until this point we have used BP bright and faint bands separately, but now we use either of these as BP depending on whether the source is brighter or fainter than G\,$=10.87$\,mag \citep{2018A&A...617A.138W}. Note that some of the sources in our catalog can have BP or RP magnitudes much fainter than 20.7\,mag in their respective bands. 
Thus in order to retrieve sources from our catalog that would have BP and RP measurements in GDR2 (or Gaia EDR3), the user may want to apply cuts to our catalog on these magnitudes.
 We do not model BP or RP excess flux spilling over from nearby sources, which will brighten up those bands for faint stars in dense areas in the Gaia data.

\subsection{Uncertainty model}\label{sec:errormodel}
In GDR2mock we used the pre-launch nominal sky-averaged error model. This underestimates uncertainties, especially in the bright regime. To simulate the Gaia measurement metrics (e.g. no.\ of observations, parallax or photometric uncertainties) more accurately for GeDR3mock, we use GDR2 data to fit a predictive model of a metric as a function of parameters that we can simulate from \texttt{galaxia} (e.g.\ magnitude, colour, position). Specifically, we select 0.5\,\% of GDR2 data at random 
and use this to train ExtraTree models \citep{Geurts2006,scikit-learn}. The first model uses Galactic longitude and latitude as inputs to predict the number of \texttt{visibility\_periods\_used} (VPU) and \texttt{phot\_g\_n\_obs} (NOBS). These are mutliplied by 34/22 (the longer baseline of Gaia EDR3 compared to GDR2) and rounded to the nearest integer. 
We then train separate models with G, BP-RP, VPU and NOBS (all values are still coming from GDR2) as inputs to predict the \texttt{parallax\_error} and \texttt{phot\_g\_mean\_mag\_error} (using approximate values computed from the symmetrised flux uncertainties). These are then scaled by $\sqrt{22/34}$ to account for the longer baseline. This scaling factor assumes that the dominant noise factor is source-noise rather than systematics, which is not actually the case at the bright end. The factor of $\sqrt{22/34}$ is the factor for flux uncertainties, not magnitude uncertainties. 
Similarly, the \texttt{radial\_velocity\_error} is predicted from G, BP-RP, and Teff and the same rescaling is applied. The procedure outlined above can be inspected in notebook~8.

From NOBS we derive NOBS for BP and RP by fitting a linear relation that minimises the least squares. 
We similarly derive the photometric uncertainties in the other bands
from linear relations on \texttt{phot\_g\_mean\_mag\_error}, and uncertainties in positions and proper motions from linear relations\footnote{The derived scaling relations compare well with the estimated end-of-mission values from \url{https://www.cosmos.esa.int/web/gaia/table-6} given the expected relative improvement of proper motion uncertainties with time.} on \texttt{parallax\_error}.
The fitted relations are listed in Table\,\ref{tab:error_scaling} and the procedure to obtain those values can be inspected in notebook~8a. We account for the $\left(22/34\right)^{1.5}$ uncertainty scaling for the proper motions. We produce the mock errors this way in order to save storage in the ADQL data base, because simple scaling relations with other columns do not require additional space.

\begin{table}[]
\caption{Empirical scaling relations that evaluate the quantity in the first column as a function of the quantity (from GDR2) in the second columns.}
    \centering
    \begin{tabular}{c|c}
     derived quantity  & scaling relation \\
       \hline
        \texttt{phot\_bp\_n\_obs}  & 0.092 \texttt{phot\_g\_n\_obs}\\
        \texttt{phot\_rp\_n\_obs}  & 0.096 \texttt{phot\_g\_n\_obs}\\
        \texttt{phot\_bp\_mean\_mag\_error}  & 19.85 \texttt{phot\_g\_mean\_mag\_error}\\
        \texttt{phot\_rp\_mean\_mag\_error}  & 9.12 \texttt{phot\_g\_mean\_mag\_error}\\
        \texttt{pmra\_error}  & 1.71 \texttt{parallax\_error}\\
        \texttt{pmdec\_error}  & 1.52 \texttt{parallax\_error}\\
        \texttt{ra\_error}  & 0.81 \texttt{parallax\_error}\\
        \texttt{dec\_error}  & 0.75 \texttt{parallax\_error}\\
    \end{tabular}
\label{tab:error_scaling}
\end{table}

\subsection{Catalog entries are reported noise-free}
\label{sec:catalogue_entries}
All quantities that we report in GeDR3mock are noise-free. Noise can be added based on the uncertainty estimates derived in Section\,\ref{sec:errormodel} from within  \texttt{ADQL}: see the example in Section\,\ref{sec:example_queries}. As in the GDR2 data model, GeDR3mock contains the phase space distribution in the following observables: {\tt ra, dec, l, b, parallax, pmra, pmdec, radial\_velocity} and similarly for the photometry, though we add an extra G$_\mathrm{RVS}$ column. A few stellar parameters have been estimated in GDR2 \citep{2018A&A...616A...8A} and these are reported together with all the other known quantities in GeDR3mock: {\tt teff\_val, ag\_val, a\_g\_val, e\_bp\_min\_rp\_val, radius\_val, lum\_val, feh, a0, initial\_mass, current\_mass, age, logg, popid, a\_bp\_val, a\_rp\_val, a\_rvs\_val}. The column descriptions can be inspected here\footnote{\url{http://dc.g-vo.org/tableinfo/gedr3mock.main}}.

\section{Selection function}\label{sec:selection}
Here we explain and investigate two effects that prevent stars from entering into the real Gaia catalog, even though they would be brighter than G=20.7\,mag and therefore inside the GeDR3mock catalog. For a proper comparison between mock and data, these selection effects should be taken into account.

\subsection{Contrast sensitivity}
\label{sec:contras_sensitivity}
When two sources in Gaia are close to each other, the fainter one might not get allocated an observational window by Gaia, depending on their separation and magnitude difference \citep{2015A&A...576A..74D}. This effect, dubbed ``contrast sensitivity'', has been quantified to some degree for GDR2 by \citet{2019A&A...621A..86B}. We used their Table\,1 to calculate, for each source in GeDR3mock, its probability to be seen, which we call ``visibility''. 
We compute and add to GeDR3mock a quantity  \texttt{d11y} that gives an integer from 0 to 100, where 0 means no visibility.
The  \texttt{ADQL} query that pre-selects close pairs and calculates \texttt{d11y} is linked to the \texttt{galaxia\_wrap} repository.  As can be seen from Table\,\ref{tab:contrast_sensitivity}, 69\,million sources have issues with too bright and too near neighbours. When accounting for the magnitude limits from GDR2 this number drops to 33\,million. GeDR3mock does not include binaries or globular clusters, which would otherwise increase those numbers.
\begin{table}[]
\caption{Number of sources in GeDR3mock with certain visibility values, for all sources (second column), and for sources brighter than the magnitude limits given in Table\,\ref{tab:mag_lim} (third column). 
}
    \centering
    \begin{tabular}{c|c|c}
      visibility & GeDR3mock & GeDR3mock with G maglim \\
       \hline
      \% & \multicolumn{2}{c}{million}\\
      \hline\hline
        0  & 34  & 16 \\
        1--50  & 20  & 10 \\
        51--99 & 15 & 7 \\
        100 & 1\,505 & 1\,304 \\
    \end{tabular}
\label{tab:contrast_sensitivity}
\end{table}

\subsection{Magnitude limit of GDR2}
\label{sec:maglim}
\begin{figure}
	\includegraphics[width=\linewidth]{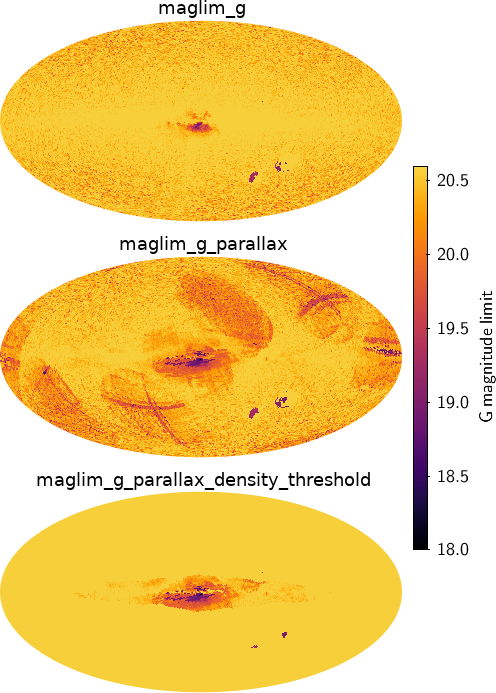}
	\caption{G magnitude limits over the sky for HEALpix level 6 in Galactic coordinates (longitude increases to the left) computed from GDR2. The colour indicates the magnitude limits. From top to bottom the maps are for sources that have: a G magnitude (all of GDR2); a parallax measurement; a parallax measurement and more than 1e5 stars per deg$^2$ -- the limit is set to 20.7\,mag in HEALpix with lower density. While for the top panel the brightest limit is 18.9\,mag, the middle and bottom panels both have one HEALpix outside the range, with 17.6\,mag.
	The name above each skymap is the column name storing the plotted magnitude; this can be queried from the auxiliary table \texttt{gedr3mock.maglim\_6}.}
	\label{fig:maglim_example}
\end{figure}

The effective magnitude limit along a line-of-sight can be shifted towards brighter magnitudes by a combination of crowding \citep{2016A&A...595A...1G} and a limited number of scans. The latter can, for faint sources, drop below the number of observations required for specific Gaia data products to be included in a release, e.g.\ parallax \citep{2018A&A...616A...2L}, G, BP, RP, or RVS. This magnitude limit can be approximated by the mode of the magnitude distribution within a specific HEALpix. When, in the following, we speak of the magnitude limit, we refer to the mode of the magnitude distribution binned in 0.1\,mag bins.
To illustrate how this manifests itself in the real data we show in Figure\,\ref{fig:maglim_example} such maps for GDR2 for G\,$<20.7$. The top panel shows the magnitude limits when only requiring G measurements. We see that in the bulge and  Magellanic Clouds the magnitude limits are brighter than everywhere else. Away from the disk the limit becomes rather noisy. In the middle panel we require that a parallax measurement be available. This makes the bulge limits brighter, and satellite scanning patterns become visible. In the bottom panel we show the same map again (requiring parallax measurement), but this time we only set the limit from the G-magnitude distribution for a HEALpix if it has more than 1e5 sources per deg$^2$. In all other HEALpix the limits are set to 20.7.

As we can see the mode estimator -- upper panel of Figure~\ref{fig:maglim_example} has two main failure modes: (a) the starcount in a specific HEALpix is low, such that the magnitude distribution gets noisy due to Poisson sampling, 
and (b) a peak in the magnitude distribution is produced by some localised stellar population in a distant overdensity, e.g.\ red clump stars in the Magellanic Clouds, that is not characteristic of the crowding limit. 
An easy fix for (a) is to only apply the magnitude limits in dense areas and
to set the magnitude limit to 20.7 in all HEALpix that have a small stellar density. This is what we do in the bottom panel of Figure\,\ref{fig:maglim_example} for a density threshold of 1e5 sources per deg$^2$\footnote{The density threshold might be useful to apply to the \texttt{maglim\_g} panel of Figure\,\ref{fig:maglim_example}. But for the \texttt{maglim\_g\_parallax} map it is obvious that there is not only low-density noise but also real structure related to the scanning law. A comparison to GDR2 would therefore be biased if we used the \texttt{maglim\_g\_parallax\_density\_threshold} map. On the other hand, we would assume that those structures related to the scanning law are suppressed in the Gaia EDR3 \texttt{maglim\_g\_parallax} map.}.

\begin{figure}
	\includegraphics[width=\linewidth]{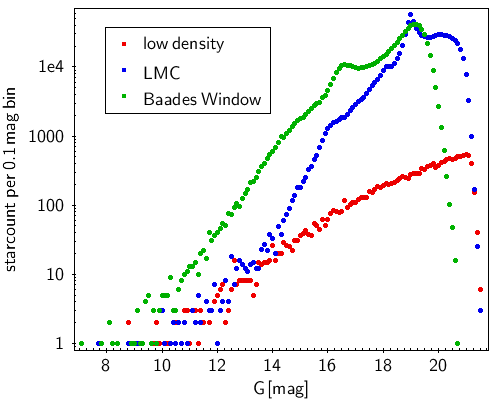}
	\caption{GDR2 G magnitude distributions in different directions of the sky. The pointings are towards Baades window, the LMC, and a low density field at l = 20, b = 30. Each curve corresponds to one HEALpix at level 6. From these magnitude distributions we approximate the limiting magnitude by taking the mode.}
	\label{fig:mag_distribution}
\end{figure}

To illustrate those failure modes further we plot in Figure\,\ref{fig:mag_distribution} the G magnitude distributions for three different HEALpix at level 6, namely towards Baades window, the LMC and a low density field at l = 20 and b = 30. The following query exemplifies the data acquisition for Figure\,\ref{fig:mag_distribution}:
\begin{lstlisting}
SELECT COUNT(*) AS ct, ROUND(phot_g_mean_mag,1) AS mag
FROM gaia.dr2light
WHERE source_id BETWEEN 4657847914607935488 AND 4657988652096290815
-- healpix level 6 pointing on Baades window
GROUP BY mag
\end{lstlisting}

We can see how the red clump peak (blue points) in the LMC at around G\,$=19$\,mag can yield an incorrect magnitude limit estimate. The low density field (red points) is not yet too noisy such that the mode of the distribution is still a good estimator for the magnitude limit but one can see how the Poisson noise in the magnitude distribution can produce modes at brighter magnitude limits if the stellar density gets even lower or the HEALpix level increases. In Baades window (green points) it can be seen that a brighter magnitude limit of about G\,=\,19\,mag is reached and sources are petering out beyond that. The red clump peak in the luminosity function of Baades window at G\,$=16$\,mag is well visible, but does not bias our magnitude limit estimation in this particular case, since the mode is at fainter magnitudes still.

We provide both variants, i.e.\ with and without a density threshold applied in the tables where the latter has no suffix and the former has \texttt{\_density\_threshold} added. We also provide magnitude limits for BP (also under the condition that G\,$<20.7$). For each band the magnitude limit comes for four flavours (and each flavour has one with and without density threshold applied):
\begin{itemize}
    \item (1) G$<20.7$\,mag (applies to all variants)
    \item (2) parallax measurement is available
    \item (3) BP and RP measurement is available
    \item (4) parallax, BP and RP are available
\end{itemize}

In Table\,\ref{tab:mag_lim} we list the number of sources that are included in the G magnitude limits for both GeDR3mock and GDR2. In total, GeDR3mock has 1,573\,M sources compared to 1,451\,M in GDR2\footnote{There are 241\,M stars in GDR2 that have G\,$>20.7$\,mag, but in this work we usually only use sources with G\,$<20.7$ unless stated otherwise.}. When considering only sources that are brighter than the HEALpix dependent magnitude limit the numbers are more similar. The reason why the mock catalog has more sources than GDR2 is because of the density limit of the Gaia instrument of about 1.05\,M sources deg$^{-2}$ \citep{2016A&A...595A...1G}. The highest density area in GeDR3mock has 5.6\,M sources deg$^{-2}$. An illustration of this can be seen in Figure\,\ref{fig:baades_window}, the CMD of Baade's window, where the magnitude limit that also requires the existence of color and parallax measurements is depicted too.

\begin{table}[]
\caption{Number of sources in GeDR3mock and GDR2 for G\,$<20.7$\,mag. Starcounts are shown for stars brighter than the limiting G magnitude given in the online table \texttt{ gedr3mock.maglim\_6}. In parenthesis the numbers are given for the stars that are outside of the magnitude limit but are still brighter than G\,$=20.7$\,mag and fulfill the selection criteria, e.g. need parallax measurement for \texttt{maglim\_g\_parallax}.}
    \centering
    \begin{tabular}{r|c|c}
     magnitude limit & GeDR3mock & GDR2 \\
       \hline
     column name  & \multicolumn{2}{c}{starcounts in million}\\
      \hline
        
   no magnitude limit & 1,573 & 1,451\phantom{ (131)} \\

        \texttt{maglim\_g} & 1,332 & 1,321 (131)\\
        \texttt{maglim\_g\_parallax} & 1,146 & 1,100 (168) \\
        \texttt{maglim\_g\_color} & 1,231 & 1,123 (131)\\
        \texttt{maglim\_g\_parallax\_color}& 1,111 & 1,012 (158)\\
        \hline
        \texttt{maglim\_g\_density\_threshold} & 1,361 & 1,358 (94)\\
    \end{tabular}
\label{tab:mag_lim}
\end{table}

We generated HEALpix maps of those magnitude limits for HEALpix levels 5, 6, and 7 (nside = 32, 64, and 128, have areas of  3.36, 0.84, and 0.21 deg$^2$ respectively) using the \texttt{gdr2\_completeness} package\footnote{\url{https://github.com/jan-rybizki/gdr2_completeness}} \citep{2018ascl.soft11018R}. They can be accessed via \texttt{gedr3mock.maglim\_X}, where X is the HEALpix level. 

Notebook 3 and 4 of \texttt{gdr2\_completeness} illustrate how to generate those maps. We encourage the user to produce maps for their specific use-cases, e.g.\ accounting for quality cuts or the existence of measurements, for example radial velocity. 

We did not provide RP magnitude limits because those are mainly governed by the condition that G is brighter than 20.7\,mag. Since RP is usually brighter than G, sources are usually lost because they get too faint in G not because they get too faint in RP. 

We also caution the use of the BP magnitude limit, because in dense areas, faint sources can acquire very bright BP (and RP) magnitudes due to flux contamination from neighbouring sources, something that is not modelled in GeDR3mock. BP maps might still be useful when comparing to other data or when modelling the BP and RP flux excess.

More details on all bands and a comparison to GeDR3mock magnitude limits can be found in appendix\,\ref{sec:app_maglim}.

Once the real data, Gaia EDR3, comes out we will provide updated magnitude limit maps in the TAP service.

An example of how to query all stars in GDR2 that are below the \texttt{maglim\_g} magnitude limit for HEALpix level 6 is given below.
\begin{lstlisting}
SELECT COUNT(*)
FROM gaia.dr2light AS g
JOIN gedr3mock.maglim_6 AS lim
ON (g.source_id/140737488355328=lim.hpx)
-- matches catalogs on HEALpix number (level 6)
WHERE phot_g_mean_mag<lim.maglim_g
-- takes about 1 to 2 hours
\end{lstlisting}

A python package with a more rigorous method providing completeness as a function of magnitude per HEALpix (Boubert \& Everall, submitted) can be found here\footnote{\url{https://github.com/DouglasBoubert/selectionfunctions}}. One drawback of this is that the magnitude limits seem to depend on the authors' all-sky partition into equal density areas.

\section{Comparison to GDR2}
\label{sec:comparison}
As a first quality assessment and to get an overview of the catalog parameters, we compare GeDR3mock with GDR2. This also serves to illustrate how the catalog can be queried via TAP services using ADQL queries. 

\subsection{Sky distribution}
In order to compare the source density over the sky between GeDR3mock and GDR2, we apply the contrast sensitivity and the magnitude limits from the previous section to GeDR3mock:
\begin{lstlisting}
SELECT COUNT(*) AS ct, hpx
FROM gedr3mock.main AS g
JOIN gedr3mock.maglim_6 AS lim
ON (g.source_id/140737488355328=lim.hpx)
WHERE phot_g_mean_mag<lim.maglim_g_density_threshold
AND d11y-RANDOM()*100 > 0
-- samples the visibility probability
GROUP BY hpx ORDER BY hpx
-- starcounts per hpx are returned
-- takes about 1 to 2 hours (varies with load)
\end{lstlisting}
This returns 1,336\,M stars.
For GDR2 this returns 1,358\,M stars.

\begin{figure}
	\includegraphics[width=\linewidth]{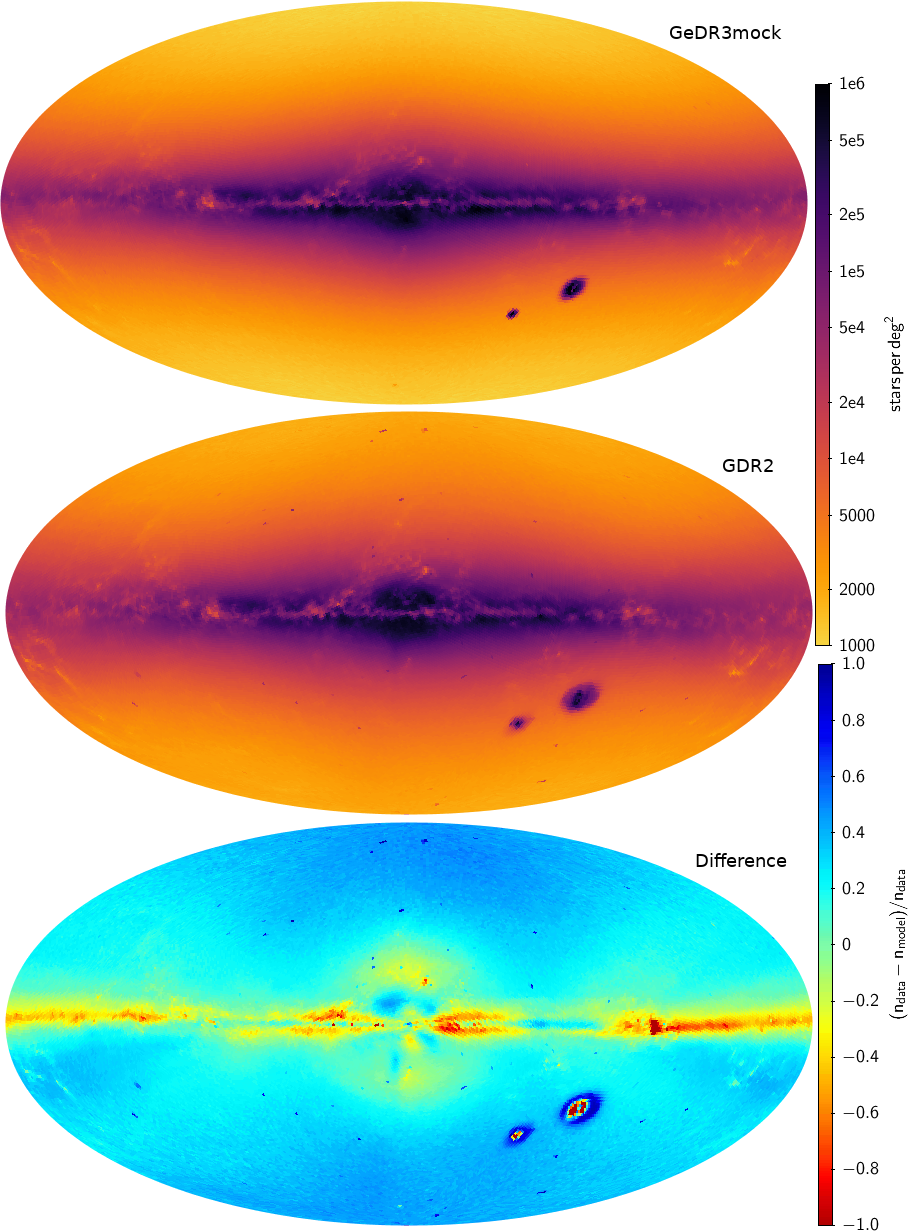}
	\caption{Stellar density in Galactic coordinates in Mollweide projection for GeDR3mock in the top, GDR2 in the middle, and the fractional differences in the bottom panel.}
	\label{fig:catalog_comparison}
\end{figure}

In Figure\,\ref{fig:catalog_comparison} we show the stellar densities in HEALpix level 6 for GeDR3mock and GDR2 and a comparison map at the bottom. GDR2 has more sources towards the poles, but overall the agreement is reasonably good. Globular clusters and the Sagittarius stream are visible, and the Magellanic Clouds show more structure in GDR2. When looking at the comparison map,
we see the footprint of the \citet{2006A&A...453..635M} extinction map transitioning into \citet{2003A&A...409..205D} (c.f. Figure\,1 of \citet{2016ApJ...818..130B}), where there is a discrete jump in the model starcounts (colour getting redder) towards the right in the galactic plane. The warp is more prominent in the model, and the bulge structure is not well reproduced. The fit to the Magellanic clouds is poor, owing to the simplistic Gaussian distribution in GeDR3mock.

\subsection{Color--magnitude diagram (CMD)}
\begin{figure*}
	\includegraphics[width=\linewidth]{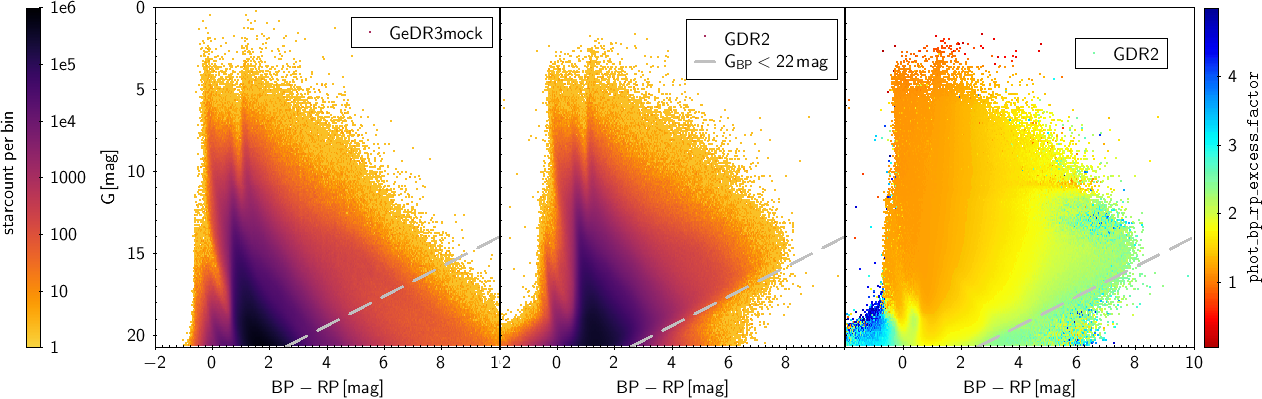}
	\caption{Colour-magnitude diagram for GeDR3mock in the left and GDR2 in the middle panel, colour-coded by number of sources per CMD bin. The right panel shows again the GDR2 CMD but this time the average  \texttt{phot\_bp\_rp\_excess\_factor} 
	per bin is depicted. The dashed grey line (its functional form is $\mathrm{G}=-0.9 (\mathrm{BP-RP})+23$) indicates a sharp limit in GeDR3mock and GDR2, below which no stars are left if we cut on G$_{\rm BP}<22$\,mag.}
	\label{fig:cmd_comparison}
\end{figure*}

Another insightful test is the CMD comparison. Here we do not apply HEALpix-dependent magnitude limits to either catalog, as those do not change the basic structure of the distributions. The query is:
\begin{lstlisting}
spSELECT COUNT(*) AS N, 
AVG(phot_bp_rp_excess_factor) AS excess,
ROUND(phot_bp_mean_mag - phot_rp_mean_mag,2) AS color, 
ROUND(phot_g_mean_mag,1) AS mag 
FROM gaia.dr2light TABLESAMPLE(50)
-- this only uses 50% of the table
WHERE phot_g_mean_mag < 20.7 
GROUP BY color, mag
-- this query takes between 1 and 2 hours
\end{lstlisting}
For GDR2 and GeDR3mock\footnote{For GeDR3mock we add measurement noise to the photometry and also half the query volume by requiring the \texttt{random\_index} to be lower than 786728660 (as TABLESAMPLE() does not work on views).}. 
These queries count the stars and average the \texttt{phot\_bp\_rp\_excess\_factor}\footnote{Excess of flux in the BP and RP integrated photometry with respect to the G band. In the absence of nearby sources this value should be close to 1. Large values indicate contamination of BP and RP photometry.} in magnitude bins (the excess factor has not been modelled in GeDR3mock). The data is shown in Figure\,\ref{fig:cmd_comparison} where the density distribution is given for GeDR3mock and GDR2 in the left and middle panel respectively. We see that GDR2 lacks sources\footnote{The term 'sources' is used for GDR2 data because not all entries are stars. For GeDR3mock the term 'star' is equivalent to 'source', since all entries are stars.} below the grey dashed line. The line indicates where the number of stars drops sharply when cutting on G$_{\rm BP}<22$\,mag. It seems to be a limit where the bulk part of sources is getting lost in GDR2 (with G$<20.7$\,mag). In the right panel of Figure\,\ref{fig:cmd_comparison} we see that sources which go below that line have issues with contaminated BP and RP measurement. Similarly the very blue stars in the GDR2 data have no counterpart in GeDR3mock. Again these stars have high \texttt{phot\_bp\_rp\_excess\_factor}, which is not modelled in GeDR3mock. The other structures in the CMD are fairly well reproduced.
With respect to catalog selection function there are only 1.6\,M sources in GDR2 (5\,M if including sources with G\,$>20.7$) with G$_{\rm BP}>22$\,mag, while GeDR3mock has 36\,M.

\section{Catalog Content \& Limitations}
\label{sec:catalog}
The catalog contains 1,573,457,319 stars. It is hosted at GAVO\footnote{\url{http://dc.g-vo.org/tap}} and can be queried via \texttt{gedr3mock.main}. Example queries see Section\,\ref{sec:example_queries}. A bulk download is also available.\footnote{\url{https://dc.zah.uni-heidelberg.de/gedr3mock/q/download/}}

\subsection{Data model and Catalog Content}
Our catalog, by design, mimicks the GDR2 data model, which will be similar in Gaia EDR3. Some fields are filled with NULLs rather than omitted in order for GDR2 ADQL queries not to throw errors. Values like \texttt{phot\_bp\_rp\_excess\_factor} or \texttt{ruwe} are not easy to model because they depend on the actual measurement, but one could train models on the real data to predict those values for the mock catalog, using the method presented in Section~\ref{sec:errormodel} (notebook 8). 

Entries in GeDR3mock that have no counterpart in the GDR2 data model are now explained:
\begin{itemize}
    \item \texttt{phot\_g\_mean\_mag\_error}\,\, For convenience we provide magnitude errors for all photometric bands. These are only good approximations of the flux error for small values. 
    \item \texttt{phot\_rvs\_mean\_mag}\,\, Since we have the isochrone models with an approximate RVS band\footnote{G$_\mathrm{RVS}$ transmission curve: \url{https://github.com/jan-rybizki/Galaxia_wrap/blob/master/notebook/isochrone_generation/passband/rvs_gedr3mock.dat}} we also provide RVS mag (simply computed assuming a Vegamag zeropoint) because it is useful to select magnitude complete RVS samples. \item \texttt{popid}\,\, The popid from the Besan\c con model (c.f. table~\ref{tab:local_mass}; halo = 8 and bulge = 9), additionally having the Magellanic clouds = 10 and the open clusters = 11.
    \item \texttt{d11y}\,\, The visibility is given in percentage. Can be lower than 100 due to bright sources in the near vicinity (see Section\,\ref{sec:contras_sensitivity}).
    \item \texttt{index\_parsec}\,\, Is an index for joining the main mock catalog
    to other photometric bands/extinctions in the \texttt{gedr3mock.parsec\_props} table.
    \item \texttt{a\_bp\_val}, \texttt{a\_rp\_val}, \texttt{a\_rvs\_val}\,\, These are extinctions in the specified bands, in analogy to \texttt{a\_g\_val} in the G band.
    \item \texttt{source\_id}\,\, The most significant bits identify the HEALpix number as with Gaia \texttt{source\_id}. The rest of the \texttt{source\_id} is a running number. The \texttt{source\_id} can be easily turned into HEALpix number for any arbitrary HEALpix level, n, smaller than or equal to 12 (level 12 corresponding to Nside = 4096) via division:
    \begin{equation}
    \mathrm{Healpix}(\mathrm{level}=n) = \text{FLOOR}\left(\frac{\mathtt{source\_id}}{2^{35}\times4^{(12-n)}}\right)
    \end{equation}
\end{itemize}
Few additional stellar parameters not listed above but can be found in Section\,\ref{sec:catalogue_entries}. General information on the catalog and its columns can be inspected here\footnote{\url{http://dc.g-vo.org/tableinfo/gedr3mock.main}}.

\subsection{Limitations}
The underlying Galaxy model is a simple approximation of reality with know shortcomings, see lower panel of Figure\,\ref{fig:catalog_comparison} and discussion thereof in Section\,\ref{sec:comparison}. There have been improvements in the thick disk, halo \citep[e.g.][]{2014A&A...569A..13R} and bulge \citep{2012A&A...538A.106R} components of the Milky Way model, but these updates did not build on each other, so we decided to stay with the basic model update from \citet{2014A&A...564A.102C}. LMC and SMC have only Gaussian distributions with inconsistent velocity prescription. We only simulate single stars. The star formation in GeDR3mock is smooth (not clumpy) and independent of the 3D extinction model, therefore the two do not show the correlations one observes in the real MW. The all-sky 3D extinction map is up-to-date but not perfect, especially where different maps have been joined together. 

\subsection{Updates when GaiaEDR3 is released}
We plan to update our mock catalog after the release of GaiaEDR3, foreseen for late 2020. This will contain magnitude limit maps as well as error, nobs, ruwe and contrast sensitivity columns based on GaiaEDR3 data. As some of those already exist, we will add abbreviations indicating that these were derived using Gaia EDR3 data. Updates to GeDR3mock will be announced here\footnote{\url{http://dc.g-vo.org/browse/gedr3mock/q}}. 
    
\subsection{Extension to GDR3 Content}
The ``full'' GaiaDR3 currently planned for late 2021 will include many more data products. To assist the use and analysis of that catalog, we plan to augment GeDR3mock in a follow-up study with:
\begin{itemize}
    \item binaries
    \item galaxies and quasars
    \item models of BPRP and RVS spectra (if publicly available)
    \item chemical abundances using chemical evolution models.
\end{itemize}

\section{Example use cases with ADQL queries}
\label{sec:example_queries}

\subsection{Distance prior}
The user might be interested in producing a distance prior for the GDR2 RVS sample to be used in a Bayesian parameter estimation similar to the distance estimation in \citet{2018AJ....156...58B} (see also \citet{2018RNAAS...2...51M}). 
Following is a query that would mimick the GDR2 RVS sample selection and returns the mean distance per HEALpix:
\begin{lstlisting}
SELECT AVG(1000/parallax) AS mean_distance,
ivo_healpix_index(5, ra, dec) AS healpix
FROM gedr3mock.main
WHERE phot_rvs_mean_mag < 12
AND teff_val < 6900
AND teff_val > 3550
-- selection mimicking RVS sample
GROUP BY healpix
-- takes about half an hour
\end{lstlisting}
The function ivo\_healpix\_index(5, ra, dec) shown here computes HEALpix indices based on RA and Dec; for Gaia and related data products, this is in general not necessary because by construction of the \verb|source_id| column one can obtain the HEALpix (in this case, of order 5) somewhat faster by computing ROUND(\texttt{source\_id}/$\left(2^{35}\times 4^{(12-5)}\right)$), but the function might be useful for tables without \verb|source_id|.

We can not use the statement ``WHERE radial\_velocity IS NOT NULL'' because in GeDR3mock all radial velocities are known. Therefore the selection function needs to be approximated.

\begin{figure}
	\includegraphics[width=\linewidth]{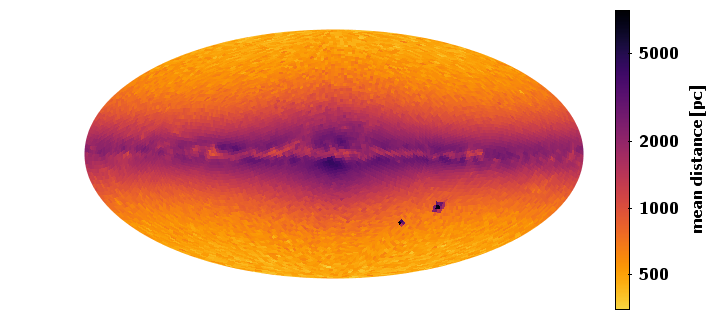}
	\caption{Mean distances over the sky in galactic coordinates in GeDR3mock with G$_\mathrm{RVS}<12$ and $3550<\texttt{teff\_val}<6900$. The color encodes mean distance logarithmically. In total this selection returns 7.1\,M sources.}
	\label{fig:rvs_sample}
\end{figure}

Figure\,\ref{fig:rvs_sample} shows the mean distance per HEALpix which could be directly used as a prior parametrisation. 7.1\,M sources are returned by GeDR3mock which is more than the 5.3\,M that GDR2 has below G$_\mathrm{RVS}=12$ (G$_\mathrm{RVS}$ needs to be approximated using Equation 2 and 3 from \citet{2018A&A...616A...1G}). The reason of course is that the effective magnitude limit is brighter in the dense parts of the sky. Cutting on G$_\mathrm{RVS}<12$ is only a first order approximation. For refinement we recommend to produce a custom magnitude limit map for the RVS sample using the \texttt{gdr2\_completeness} package.

\subsection{Parallax uncertainty}
\begin{figure*}
	\includegraphics[width=\linewidth]{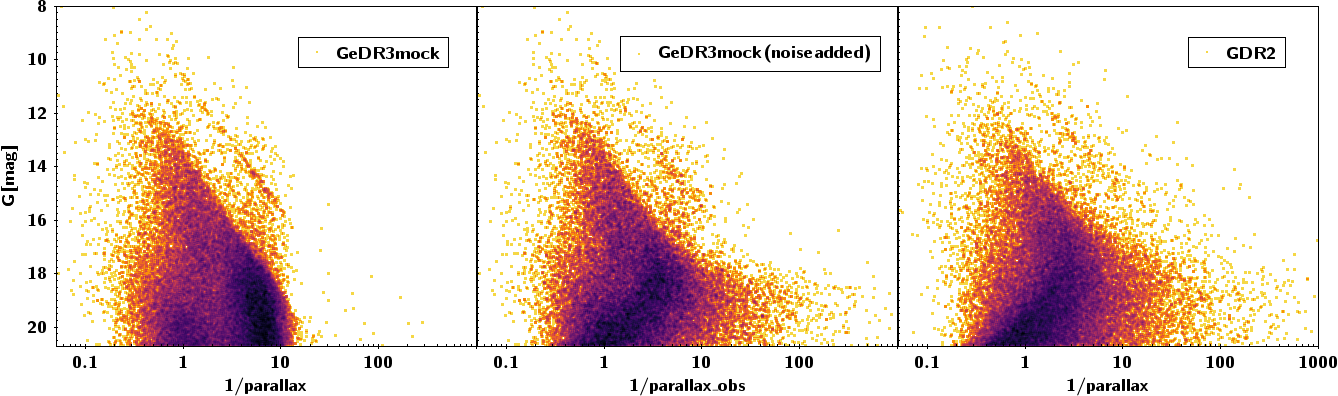}
	\caption{Inverted parallaxes (in mas) vs. G magnitude for GeDR3mock, GeDR3mock noise added and GDR2 from left to right.}
	\label{fig:parallax_error}
\end{figure*}

Because measured parallaxes can have very large uncertainties, the distribution of measured parallaxes can be quite different than for mock (true) parallaxes. We show this for the HEALpix 7876 (at level 5) which is a low density out-off-plane field at l = 20, b = 30. GDR2 contains 46\,k sources (G\,$<20.7$) and GeDR3mock has 39\,k sources in that HEALpix.  
Figure\,\ref{fig:parallax_error} shows, from left to right, the inverted parallax vs the G for: GeDR3mock; the same with parallax noise added; GDR2. In the absence of measurement uncertainty on the left we see a bimodal distribution in parallax at G=20.7, the peak at 1\,kpc consists mainly of lower main sequence stars while the one at about 8\,kpc consists mainly of upper main sequence and turn-off stars. These two sequences merge when the parallax uncertainty is added (the G magnitude error is negligible in this diagram). Similarly, the diagonal line in the top right of the three CMDs, which corresponds to the red clump, becomes blurred once noise is added. Noise can be added from within ADQL using:
\begin{lstlisting}
SELECT parallax, phot_g_mean_mag,
GAVO_NORMAL_RANDOM(parallax,parallax_error) AS parallax_obs
FROM gedr3mock.main
WHERE source_id BETWEEN 4433793833146253312 AND 4434356783099674623
-- only a low-density HEALpix of level 5
-- takes few seconds
\end{lstlisting}
The numbers in the ``WHERE'' statement are $2^{35}\times 4^{(12-5)}\times7876$ and $2^{35}\times 4^{(12-5)}\times7877-1$. The above ADQL query produces the data for the plot together with the analog query for GDR2. 

\subsection{CMD in Baade's window - magnitude limit}
\begin{figure*}
	\includegraphics[width=\linewidth]{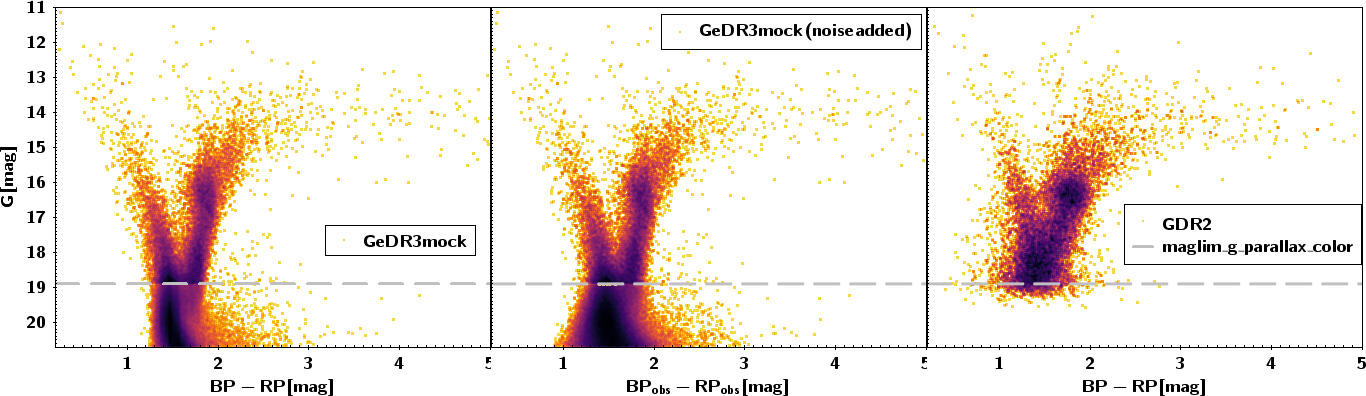}
	\caption{CMD of Baade's Window (a circle of 0.1\,degree) for GeDR3mock, GeDR3mock noise added, and GDR2 (panels from left to right). The empirically determined g magnitude limit is indicated as a grey line. The density distribution is renormalised above the magnitude limit.}
	\label{fig:baades_window}
\end{figure*}

Here we look at a CMD in a high density area, namely Baade's window.
This time we add noise to the mock photometry and compare to GDR2 data. The G magnitude limit, when parallax and BP and RP measurement are required, is 18.9\,mag. We only query in a circle of radius 0.1\,degree to keep the runtime short (query runs in synchronous mode). GDR2 contains 13\,k sources, whereas GeDR3mock contains 134\,k sources. When applying the magnitude cut, these numbers change to 12\,k and 18\,k respectively. The GeDR3mock data for Figure\,\ref{fig:baades_window} comes from the following query:

\begin{lstlisting}
SELECT phot_g_mean_mag, phot_bp_mean_mag, phot_rp_mean_mag,
GAVO_NORMAL_RANDOM(phot_g_mean_mag,phot_g_mean_mag_error) AS g_obs,
GAVO_NORMAL_RANDOM(phot_bp_mean_mag,phot_bp_mean_mag_error) AS bp_obs,
GAVO_NORMAL_RANDOM(phot_rp_mean_mag,phot_rp_mean_mag_error) AS rp_obs
FROM gedr3mock.main
WHERE DISTANCE(270.879, -30.022, ra, dec)< 0.1
-- takes a few seconds
\end{lstlisting}
The left (blue) plume contains upper main sequence and turn-off stars, while the right (red) plume contains giant stars. The overdensity at G\,=\,16\,mag is the red clump (c.f. Figure\,\ref{fig:mag_distribution}). Both plumes seem to have merged at fainter magnitudes in GDR2, whereas even with noise applied these remain distinct in mock. Only at fainter magnitudes does the noise become visible, as seen in the spreading in colour.

\subsection{Local 50\,pc sample}
The local normalisation, i.e.\ the local stellar mass density, is a common benchmark for Galaxy models, we query the 50\,pc sample using:
\begin{lstlisting}
SELECT initial_mass, current_mass, age, popid, feh, bp_rp, phot_g_mean_mag, parallax, 
GAVO_NORMAL_RANDOM(parallax,parallax_error) AS parallax_obs
FROM gedr3mock.main
WHERE 1/parallax < 0.05
-- takes about 10 minutes
\end{lstlisting}
This returns 49,934 sources.

\begin{figure}
	\includegraphics[width=\linewidth]{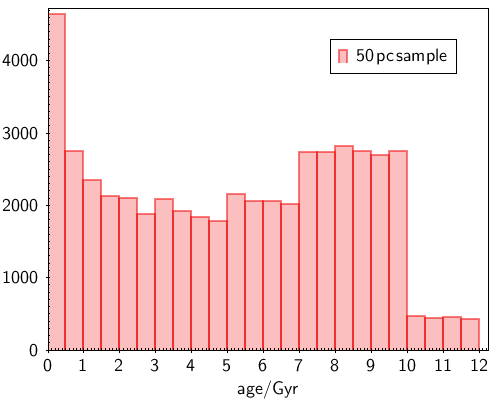}
	\caption{The age distribution of the 50\,pc sample. }
	\label{fig:local_age}
\end{figure}

\begin{figure}
	\includegraphics[width=\linewidth]{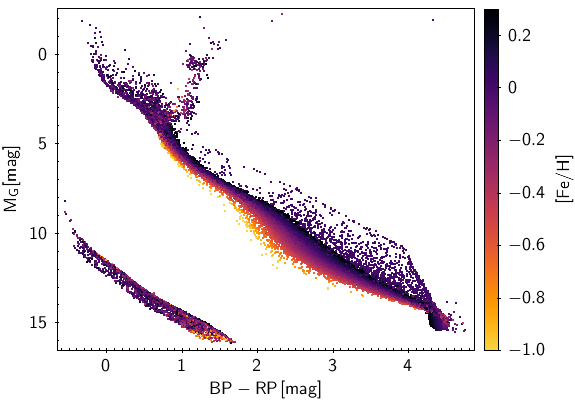}
	\caption{The colour absolute magnitude diagram of the 50\,pc sample, with metallicity colour coded. In total we have 50\,k sources with 4\,k being white dwarfs.}
	\label{fig:local_cmd}
\end{figure}

In Table\,\ref{tab:local_mass} we compare our local mass density to Model B of \citet{2014A&A...564A.102C}, their table 7. The mass values agree pretty well, just the thick disk is only about 5\% of the local mass density, compared to their 9\%. 
\begin{table}[]
\caption{Contribution of all galactic components to the local stellar mass density. We compare to Model B from \citet{2014A&A...564A.102C}.}
    \centering
    \begin{tabular}{c|c|c|c}
     popid & age [Gyr]  & GeDR3mock & Model B \\
       \hline
       \phantom{1} & [Gyr] & \multicolumn{2}{c}{10$^{-3}\times$M$_\odot$\,pc$^{-3}$}\\
       \hline
Thin disk 0& 0 - 0.15 & 1.7 & 1.9 \\
    \phantom{Thin disk }1 & 0.15 - 1  & 4.9 & 5.0\\
        \phantom{Thin disk }2 & 1 - 2     & 3.6 & 4.1\\
        \phantom{Thin disk }3 & 2 - 3     & 3.1 & 2.8\\
        \phantom{Thin disk }4 & 3 - 5     & 5.4 & 4.9\\
        \phantom{Thin disk }5 & 5 - 7     & 5.7 & 5.0\\
        \phantom{Thin disk }6 & 7 - 10    & 11.1& 9.3\\
        Total thin disk & 0 - 10 & 35.5 & 33.0 \\
        White dwarfs & 0 - 12 & 5.0 & 7.1\\
        Thick disk 7 & 10 - 12& 1.7 & 2.9\\
    \end{tabular}
\label{tab:local_mass}
\end{table}
Figure\,\ref{fig:local_age} shows the age distribution of the local 50\,pc sample. The piecewise flat, but exponentially decreasing SFR (for thin-disk \texttt{popid} 0 to 6) is visible, as well as a local overdensity of very young (dynamically cold) stars.
All stars of the 50\,pc sample are depicted in Figure\,\ref{fig:local_cmd} together with their respective metallicities (colour coded). The sample contains 4,162 white dwarfs (WD) for which the current mass is much lower than the initial mass. Extrapolating to a 100\,pc sample 10,856 of these WDs would be within the completeness range of \citet{2018MNRAS.480.4505J}. They find 8,555 stars, which is only a 20\,\% difference. 

The stellar distribution in the CMD looks reasonably well but the pre-main sequence might be a bit too pronounced, as it was in GDR2mock \citep{2018PASP..130g4101R}. 
We find that 214 (0.4\%) mainly faint sources would have scattered out of our 50\,pc sample if cutting on observed parallax. Vice versa 236 that are truly outside of 50\,pc would have scattered in when cutting on observed parallax. The 10\,\% increase for the in-scattering stars is due to the assymetric volume at the border of the 50\,pc sphere, given that the stellar density is almost isotropic.

\subsection{Other photometric bands}
\label{sec:parsec_props}
It is possible to query the absolute magnitudes and extinctions in other bands (UBV, 2MASS, SDSS) for specific values of $A_0$ via the \texttt{gedr3mock.parsec\_props} table. An example query for apparent magnitudes in 2MASS bands could be:
\begin{lstlisting}
SELECT phot_g_mean_mag, bp_rp AS color, tmass_j-5*LOG10(parallax/100)+a0*A0_1_tmass_j AS tmass_j, 
tmass_ks-5*LOG10(parallax/100)+a0*A0_1_tmass_ks AS tmass_ks 
FROM gedr3mock.main 
JOIN gedr3mock.parsec_props 
USING (index_parsec)
-- crossmatching with the PARSEC isochrone table
WHERE source_id BETWEEN 4433793833146253312 AND 4434356783099674623
-- takes a few seconds
\end{lstlisting}
Beware that all values in the \texttt{parsec\_props} table are binned according to the procedure outlined in Section\,\ref{sec:isochrones} (they are mapped onto the catalog stars via \texttt{index\_parsec}), which means that the CMD distribution will be somewhat discretised. Also by using \texttt{A0\_1\_tmass\_j} in the above query we approximate the extinction with a low $A_0$ value, which means that for large $A_0$ values the extinction will be overestimated since the absorption does not scale linearly with $A_0$ as the source spectrum gets redder with dust column density. 


\section{Summary}
\label{sec:summary}
We have presented the generation and content of a Gaia early DR3 mock stellar catalog (GeDR3mock). With respect to the previous version, GDR2mock \citep{2018PASP..130g4101R}, we have updated the thin disk model \citep{2014A&A...564A.102C} as well as the 3d extinction map \citep{2019arXiv190502734G} and isochrones \citep{2017ApJ...835...77M}, which now also include white dwarfs \citep{Bertolami_2016}. We also added a simple model of the Magellanic Clouds and open clusters, the latter including internal rotation.

We refined the uncertainty model by training empirically on GDR2 data and scaled it to the longer time baseline of Gaia EDR3. A main focus of our investigation is modelling the selection function of the Gaia instrument and DPAC filtering. We provide all-sky magnitude limit maps \citep{2018ascl.soft11018R}  approximated empirically by the mode of the magnitude distribution in a specific line-of-sight. A better comparison between model and data is achieved when applying those cuts to the relevant subsets. Similarly we investigate how many sources in GeDR3mock would suffer from decreased visibility due to contrast sensitivity\,\citep{2019A&A...621A..86B} and flag those stars in GeDR3mock.

In order for the user to be able to create their own synthetic stellar catalog from N-body data or a galaxy model, we provide the routines we used for generating our catalog in the python package \texttt{galaxia\_wrap} \citep{2019ascl.soft01005R}, as well as the isochrones and the modified \texttt{galaxia} software, and the jupyter notebooks that illustrate their use\footnote{\url{https://github.com/jan-rybizki/Galaxia_wrap}}.

We provided some example \texttt{ADQL} queries to show the many possible catalog interactions and to compare GDR2 to our mock stellar catalog. We plan to add columns/tables that update the magnitude limits and uncertainty estimates once Gaia EDR3 is released. These additions will be announced on the GAVO site of the catalog\footnote{\url{http://dc.g-vo.org/browse/gedr3mock/q}}. In preparation for the "full" Gaia DR3, we plan to augment GeDR3mock with data products that will be new in full Gaia DR3, including binaries, extragalactic objects, and chemical abundances.

\section{Acknowledgements}

This work made use of the following software packages: \texttt{topcat} \citep{2005ASPC..347...29T}, \texttt{HEALpix} \citep{2005ApJ...622..759G}, \texttt{astropy} \citep{2018AJ....156..123A}, \texttt{mwdust} \citep{2016ApJ...818..130B}, \texttt{dustmaps} \citep{2018JOSS....3..695M}, \texttt{amuse} \citep{2009NewA...14..369P}.

We estimate the CO$_2$ footprint of this publication as follows: 6
person-months of work (MPIA yearly average per employee: 9 tons) = 4.5 tons. Data access: 3\,years * 1\,KW (conservative estimate of server electricity consumption) * 5\% (GeDR3mock consumed data volume) = 1.3\,MWh corresponding to 0.6\,tons CO$_2$ with the average German energy mix. I will not travel anywhere by plane for the purpose of promoting this paper.

We thank the anonymous referee for their thorough inspection and helpful comments.

We thank the German Astrophysical Virtual Observatory\footnote{\url{http://www.g-vo.org/}} for the publishing platform and for fruitful discussions on the technical
aspects of this endeavor. 

YC acknowledges support from the ERC Consolidator Grant funding scheme (project STARKEY, G.A. n. 615604).

This research has made use of the VizieR catalog access tool, CDS, Strasbourg, France (DOI: 10.26093/cds/vizier). The original description of the VizieR service was published in A\&AS 143, 23

This work has made use of data from the European Space Agency (ESA) mission Gaia, processed by the Gaia Data Processing and Analysis Consortium (DPAC). Funding for the DPAC has been provided by national institutions, in particular the institutions participating in the Gaia Multilateral Agreement. 

This work was supported by the MINECO (Spanish Ministry of Economy) through grant ESP2016-80079-C2-1-R and RTI2018-095076-B-C21 (MINECO/FEDER, UE), and MDM-2014-0369 of ICCUB (Unidad de Excelencia 'María de Maeztu'). TCG acknowledges support from Juan de la Cierva - Formaci\'on 2015 grant, MINECO (FEDER/UE).
This work was funded by the DLR (German space agency) via grant 50\,QG\,1403.
 
\bibliographystyle{aasjournal} 
\bibliography{adslib,otherlib} 

\appendix

\section{Magnitude limits derived from GDR2 and GeDR3mock}
\label{sec:app_maglim}
Here we show the magnitude limit maps for G, BP and RP derived as explained in Section\,\ref{sec:maglim} at HEALpix level 7 for GeDR3mock and GDR2. For GDR2 we required: a parallax measurement, a color measurement, and G\,$<20.7$ (which corresponds to \texttt{gedr3mock.maglim\_7.maglim\_g\_parallax\_color}). For GeDR3mock we required nothing, which implies G\,$<20.7$.

Starting with the G band, we see that in GeDR3mock the magnitude limit is generally G\,=\,20.7 which is expected from the way the catalog is generated. Exceptions from this are the central parts of the LMC and low density areas towards the Galactic poles. The former is due to the red clump producing a peaked luminosity function at the distance modulus of the LMC and the latter is due to Poisson noise in the magnitude distribution in low density HEALpix. For GDR2 we also see magnitude limits below G\,=\,20.7 in low density areas and in the Magellanic clouds, albeit a different pattern to the GeDR3mock LMC. Additionally we see values as low as G\,=\,15\,mag in the bulge, but also few scanning patterns which come exclusively from the parallax measurement requirement.

\begin{figure*}
	\includegraphics[width=\linewidth]{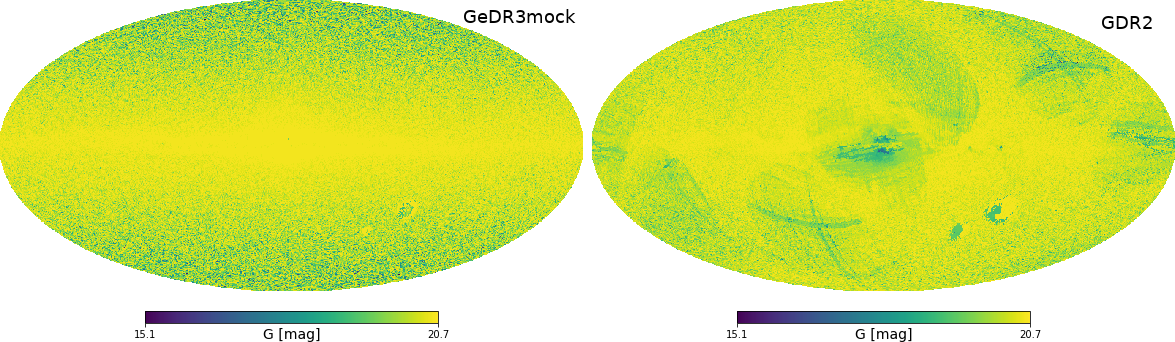}
	\caption{Magnitude limit in G for GeDR3mock and GDR2 catalogs. For GDR2 it was required to have parallax and colour measurement and to have G$<20.7$. The projection is in Galactic coordinates and in HEALpix level 7. The colorbar is the same in both panels and the ranges are the limits for GDR2.}
	\label{fig:g_maglim}
\end{figure*}

For the BP band, we have to remember, that we conditioned our queries on G\,$<20.7$\,mag and also that BP is usually fainter than G. Therefore the magnitude limits of BP can be fainter than for G which is apparent in the disk for GeDR3mock. This time the outskirts of the LMC and also the SMC have bright magnitude limits. GDR2 looks similar with respect to the MCs and also has quite faint limits in the disks high extinction areas but again the bulge and scanning law patterns have brighter magnitude limits. We have to also keep in mind that in high density areas sources in GDR2 experience BP and RP flux excess. Which can also drive the magnitude limits to the brighter end.

\begin{figure*}
	\includegraphics[width=\linewidth]{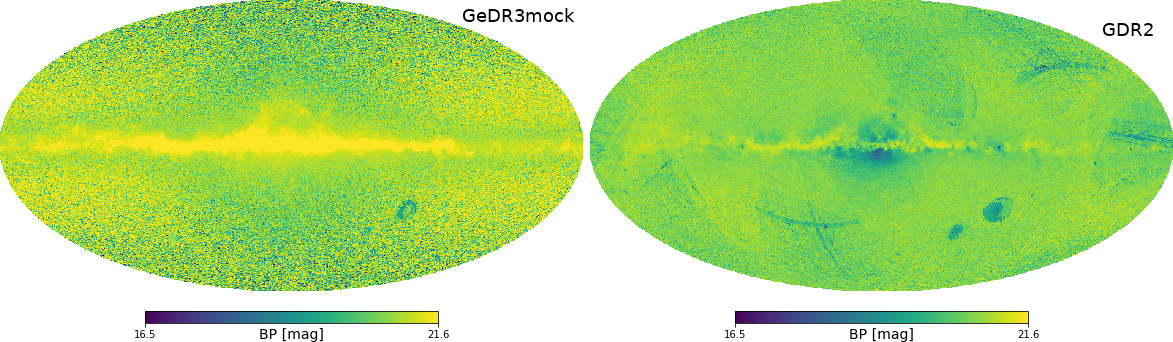}
	\caption{Same as Figure\,\ref{fig:g_maglim} but this time for BP magnitude, beware of the different colorbar limits.}
	\label{fig:bp_maglim}
\end{figure*}

RP for most stars is usually brighter than G. Therefore the magnitude limits are somewhat brighter as well, since we condition on G\,$<20.7$. Again the LMC sticks out in GeDR3mock footprint but this time we also see a small band in the high extinction areas in the mock catalog. In the high extinction areas the G limit of 20.7 cuts out fainter sources, that would have still be seen in RP but did not make it into the catalog. In the real data of GDR2 this effect can also be seen. As well as the usual bulge and scanning law pattern. The SMC and LMC can also be picked up. While the LMC has a highly lopsided feature that can also be picked up in the G and BP magnitude limit maps.

\begin{figure*}
	\includegraphics[width=\linewidth]{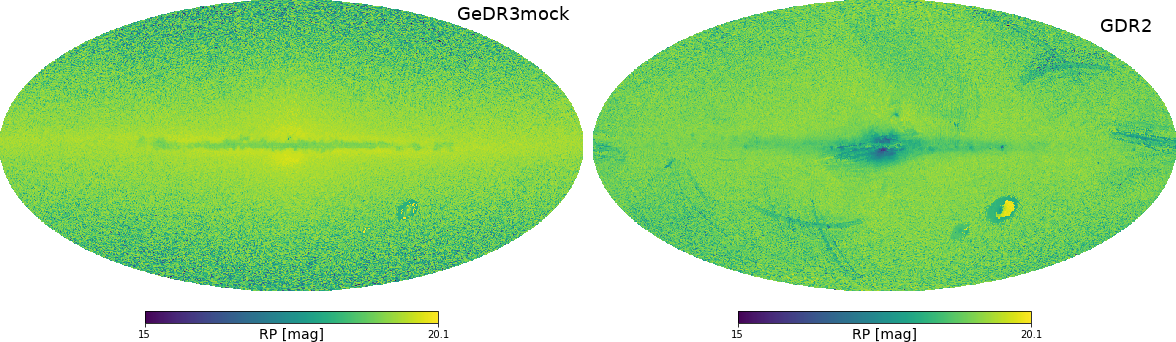}
	\caption{Same as Figure\,\ref{fig:g_maglim} but this time for RP magnitude, beware of the different colorbar limits.}
	\label{fig:rp_maglim}
\end{figure*}




\section{popid queries}
\begin{figure*}
	\includegraphics[width=\linewidth]{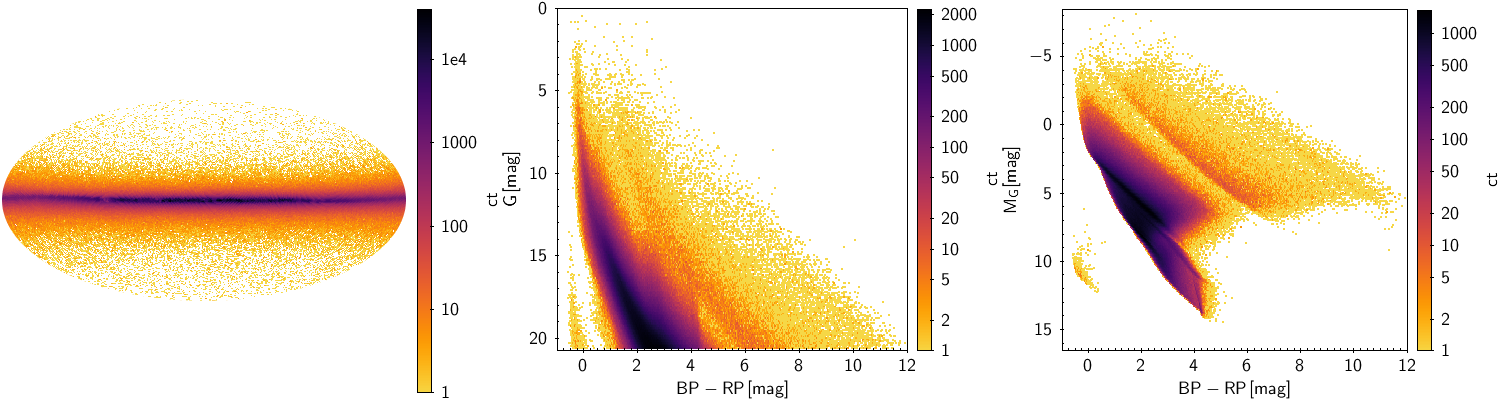}
	\caption{Overview of young thin disk (popid=0). Left an all-sky view in galactic coordinates and aitoff projection of starcounts per HEALpix in level 6. Middle and right panels are the CMD and the CMD with reddened abolute magnitudes, also colored by sources per bin.}
	\label{fig:popid0}
\end{figure*}

As a supplement to the general overview of the GeDR3mock we show here three different plots for each of the following populations:
\begin{itemize}
    \item 0 = young thin disk
    \item 1 - 6 = thin disk
    \item 7 = thick disk
    \item 8 = halo
    \item 9 = bulge
    \item 10 = Magellanic clouds
    \item 11 = open cluster
\end{itemize}
In the Figures\,\ref{fig:popid0} following we show from left to right: All-sky stellar density distribution, a binned CMD coloured by number of sources and a binned CMD using reddened absolute magnitudes also coloured by number of sources. The following queries led to the data for these figures, respectively:  
\begin{lstlisting}
-- All sky map
SELECT Count(*) AS ct,
ivo_healpix_index(6, ra, dec) AS hpx
FROM gedr3mock.main
WHERE popid = 0
GROUP BY hpx ORDER BY hpx
-- duration depends on population
\end{lstlisting}
\begin{lstlisting}
-- CMD
SELECT COUNT(*) AS ct,
ROUND(phot_bp_mean_mag - phot_rp_mean_mag,2) AS
color,
ROUND(phot_g_mean_mag,1) AS mag
FROM gedr3mock.main
WHERE popid = 0
GROUP BY color, mag
-- duration depends on population
\end{lstlisting}
\begin{lstlisting}
-- CMD using reddened absolute magnitudes
SELECT COUNT(*) AS ct,
ROUND(phot_bp_mean_mag - phot_rp_mean_mag,2) AS
color,
ROUND(phot_g_mean_mag + 5*log10(parallax/100),1) AS mag
FROM gedr3mock.main
WHERE popid = 0
GROUP BY color, mag
-- duration depends on population
\end{lstlisting}

\begin{figure*}
	\includegraphics[width=\linewidth]{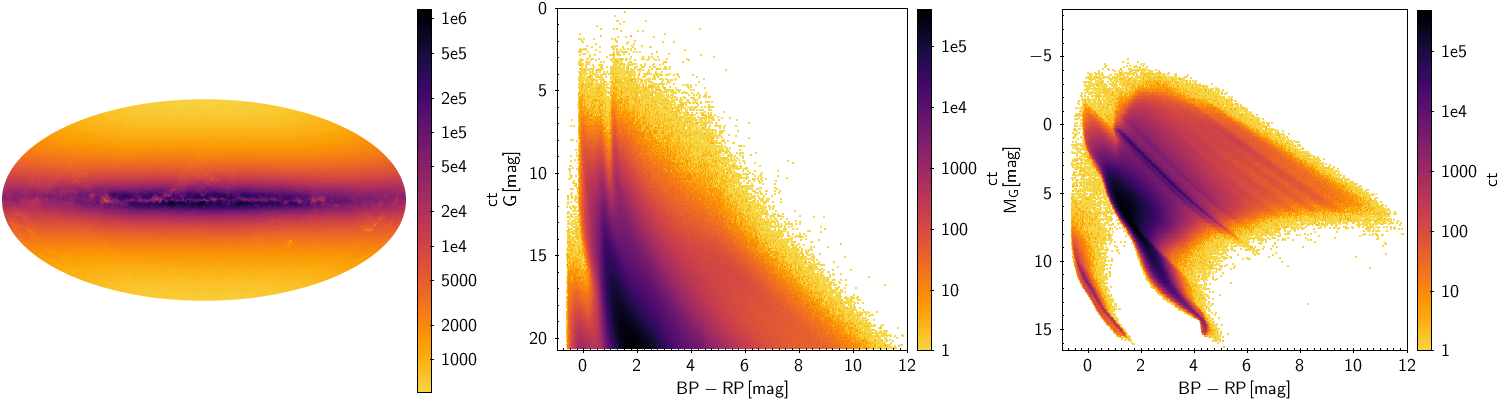}
	\caption{As Figure\,\ref{fig:popid0} but for the remaining thin disk, i.e. 0 $<$ \texttt{popid} $<$ 7.}
	\label{fig:popid16}
\end{figure*}

\begin{figure*}
	\includegraphics[width=\linewidth]{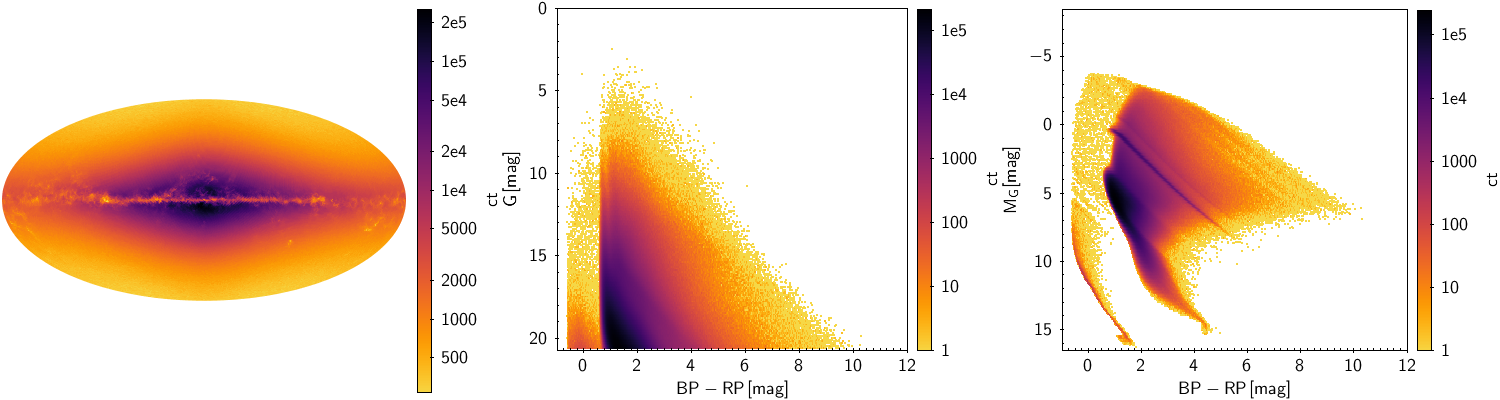}
	\caption{As Figure\,\ref{fig:popid0} but for the thick disk, i.e. \texttt{popid} = 7.}
	\label{fig:popid7}
\end{figure*}

\begin{figure*}
	\includegraphics[width=\linewidth]{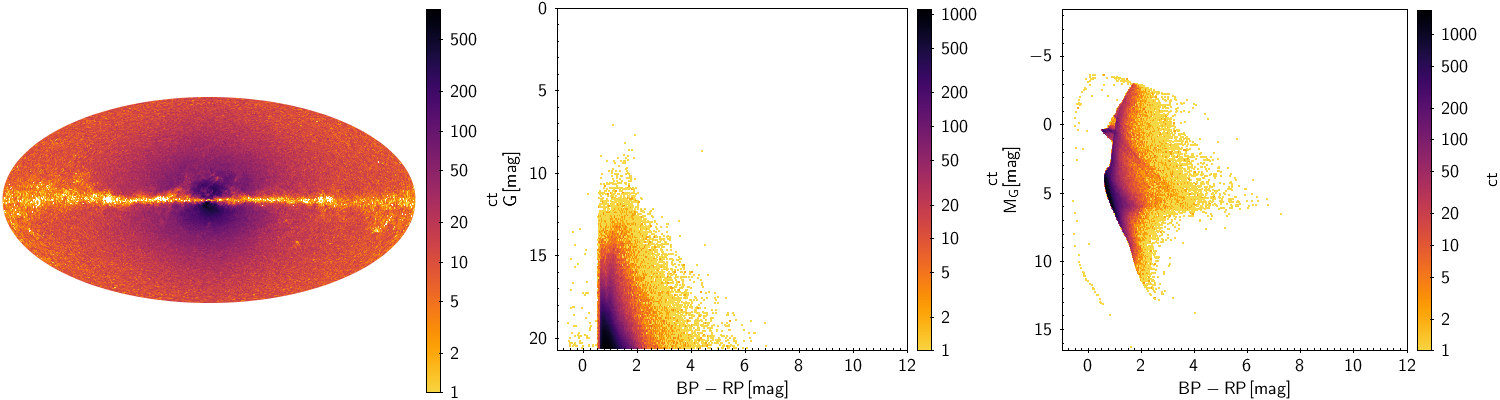}
	\caption{As Figure\,\ref{fig:popid0} but for the halo, i.e. \texttt{popid} = 8.}
	\label{fig:popid8}
\end{figure*}

\begin{figure*}
	\includegraphics[width=\linewidth]{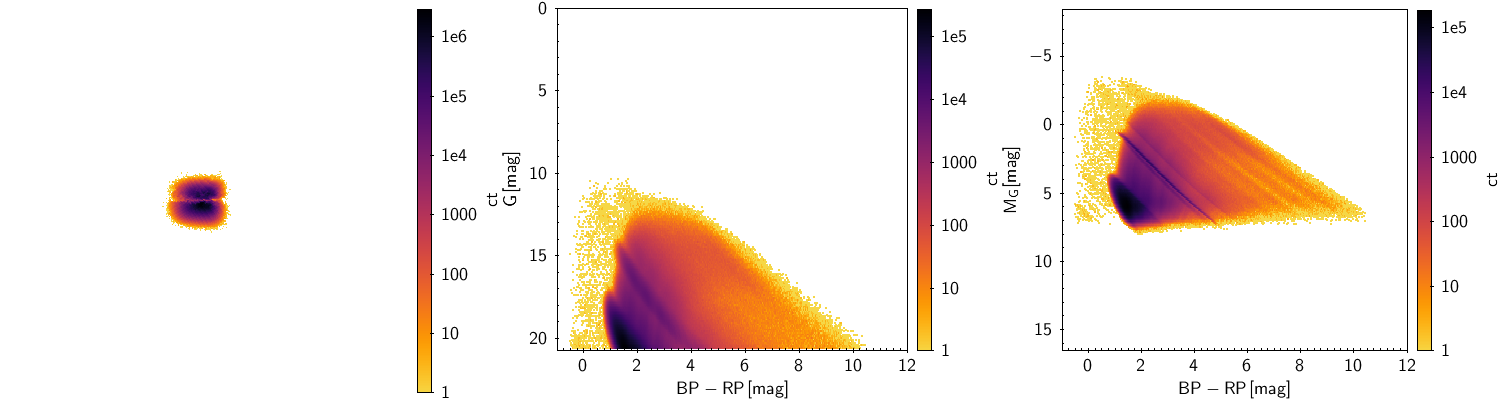}
	\caption{As Figure\,\ref{fig:popid0} but for the bulge, i.e. \texttt{popid} = 9.}
	\label{fig:popid9}
\end{figure*}

\begin{figure*}
	\includegraphics[width=\linewidth]{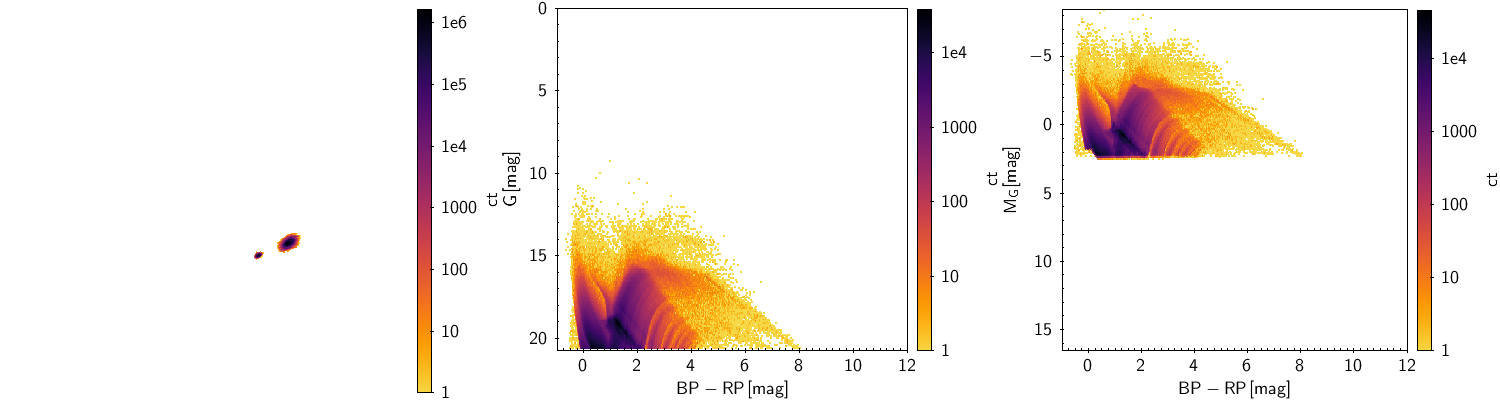}
	\caption{As Figure\,\ref{fig:popid0} but for the Magellanic clouds, i.e. \texttt{popid} = 10.}
	\label{fig:popid10}
\end{figure*}

\begin{figure*}
	\includegraphics[width=\linewidth]{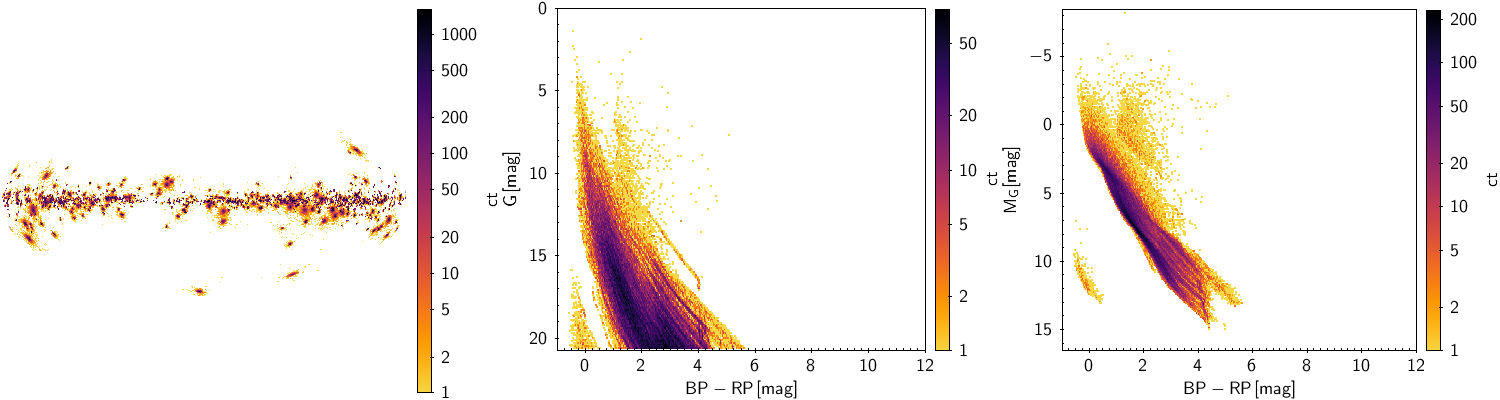}
	\caption{As Figure\,\ref{fig:popid0} but for the open cluster, i.e. \texttt{popid} = 11.}
	\label{fig:popid11}
\end{figure*}

\end{document}